\documentclass[twocolumn,aps,prl,superscriptaddress]{revtex4-2}
\usepackage{amsmath}
\usepackage{graphicx}
\usepackage{MnSymbol}
\usepackage{hyperref}
\usepackage{fancyhdr}
\usepackage{ragged2e}

\begin{document}

\title{Air, helium and water leakage in rubber O-ring seals with application to syringes}

\author{C. Huon}
\affiliation{Sanofi, 13, quai Jules Guesde-BP 14-94403 VITRY SUR SEINE Cedex, France}
\author{A. Tiwari}
\affiliation{PGI-1, FZ J\"ulich, Germany}
\affiliation{MultiscaleConsulting, Wolfshovener str. 2, 52428 J\"ulich, Germany}
\author{C. Rotella}
\affiliation{Sanofi, 13, quai Jules Guesde-BP 14-94403 VITRY SUR SEINE Cedex, France}
\author{P. Mangiagalli}
\affiliation{Sanofi, 13, quai Jules Guesde-BP 14-94403 VITRY SUR SEINE Cedex, France}
\author{B.N.J. Persson}
\affiliation{PGI-1, FZ J\"ulich, Germany}
\affiliation{MultiscaleConsulting, Wolfshovener str. 2, 52428 J\"ulich, Germany}

\begin{abstract}
\centering{\bf ABSTRACT}\\
\justifying
We study the leakage of fluids (liquids or gases) in syringes with glass barrel,  
steel plunger and rubber O-ring stopper. The leakrate
depends on the interfacial surface roughness and on the viscoelastic properties of the rubber.
Random surface roughness is produced by sandblasting the rubber O-rings.
We present a very simple theory for gas flow which takes into account both the diffusive and ballistic 
flow. The theory shows that the interfacial fluid flow (leakage) channels
are so narrow that the gas flow is mainly ballistic (the so called Knudsen limit). 
We compare the leakrate obtained using air and helium. For barrels filled with water we observe no leakage even
if leakage occurs for gases. We interpret this as resulting from capillary (Laplace pressure 
or surface energy) effects.
\end{abstract}

\maketitle

\thispagestyle{fancy}

{\bf 1. Introduction}

All solids have surface roughness which has a huge influence on a large number of physical
phenomena such as adhesion, friction, contact mechanics and the leakage of seals\cite{Ref1,Ref2,Ref3,Ref4,Ref5,Ref6,Ref7,Ref8}.
Thus when two solids with nominally flat surfaces are squeezed into contact, unless the applied
squeezing pressure is high enough, or the elastic modulus of at least one of the solids low enough,
a non-contact region will occur at the interface. If the non-contact region percolate
open flow channels exist, extending from one side of the nominal contact region to the other side.
This will allow fluid to flow at the interface from a high fluid pressure 
region to a low pressure region. 

For elastic solids with randomly rough surfaces the contact area percolate when the 
relative contact area $A/A_0 \approx 0.42$ (see \cite{Dapp}), where $A_0$ is the nominal 
contact area and $A$ the area of real contact (projected on the $xy$-plane). When the
contact area percolate there is no open (non-contact) channel at the interface 
extending across the nominal contact region,
and no fluid can flow between the two sides of the nominal contact.

The discussion above is fundamental for the leakage of static seals. Here we are interested
in rubber seals, e.g., rubber O-ring seals, or rubber stoppers for
syringes. In the latter application, the contact between the ribs 
on the rubber stopper and the (glass or polymer) barrel must be so tight that no or negligible
fluid can flow from inside the syringe to the outside. In addition,
container closure integrity is very important 
so that no microorganism can penetrate from the outside to
inside the syringe. Since the smallest microorganism (prion) may be only $\approx 10 \ {\rm nm}$
in diameter (the smallest virus is a few times larger), it is clear that 
complete container closure integrity would imply
that the most narrow junction (denoted critical junction) in the largest open (non-contact)
interfacial channel should be at most $10 \ {\rm nm}$. When this condition is satisfied, the
fluid leakage is also negligible.   

\begin{figure}
        \includegraphics[width=0.49\textwidth]{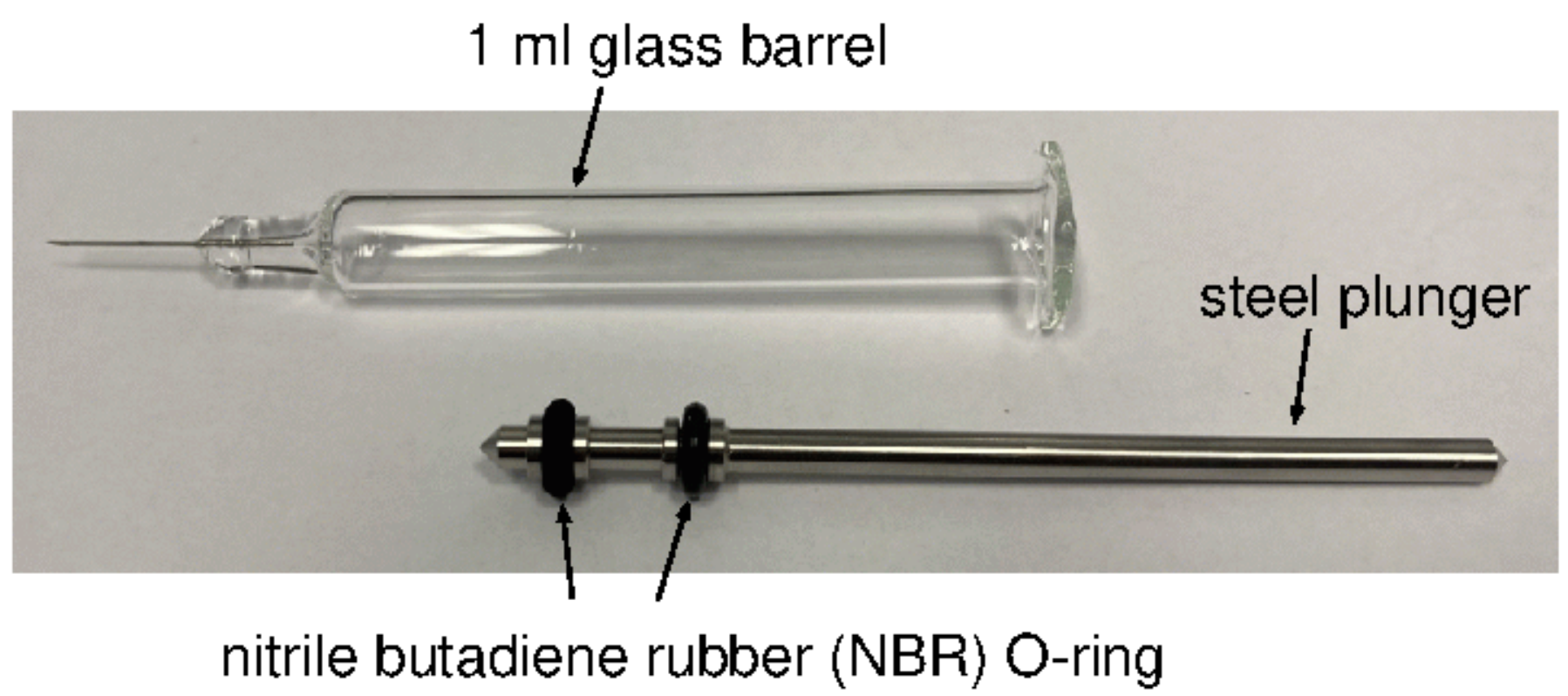}
        \caption{\label{COMBINED1.eps}
A syringe used in the present experiments consisting of a 1 ml 
glass barrel and a steel plunger with rubber O-rings.
The O-rings are sandblasted to give well defined random surface roughness.
}
\end{figure}

We have prepared model syringes to study the leakage of fluids (liquids or gases) in rubber seals. 
We have used glass barrels and steel plunger with two rubber O-rings as the stopper, see Fig.
\ref{COMBINED1.eps}. The O-rings
are sandblasted to give well defined random surface roughness. The surface roughness of the O-rings is measured using
a stylus instrument and the surface roughness power spectra is calculated. The He gas, air and water leakage
rates are measured for O-rings with different surface roughness.
The leakrate depends on the interfacial surface roughness, but only weakly
on the viscoelastic properties of the rubber.
We show that the surface separation at the most narrow constrictions 
along the percolating interfacial fluid flow (leakage) channels
are typically below $100 \ {\rm nm}$. This is smaller than the (average) gas molecule mean-free-path (due to
collisions between the gas molecules), which imply that the gas flow between the collisions 
with the solid walls is mainly ballistic rather than diffusive
(the so called Knudsen limit). We present a very simple theory for the ballistic gas 
flow which is used in combination with the Persson contact mechanics theory to predict the 
helium (He), air and water leakrate. The calculated leakrates show good correlation with the measured
leakrates.

\vskip 0.5cm
{\bf 2 Leakrate theory for liquids}

In calculating the fluid (here air, He or water) leakrate we have used the effective medium approach combined
with the Persson contact mechanics theory for the probability distribution of surface separations.
The most important region for the sealing is a narrow strip at the center of the nominal contact region,
where the contact pressure is highest (and the surface separation smallest), but the study presented below
takes into account the full pressure profile $p(x)$.

The basic contact mechanics picture which can be used to estimate the leak-rate
of seals is as follows: Consider first a seal where the nominal contact area
is a square. The seal separate a high-pressure fluid
on one side from a low pressure fluid on the other side, with the pressure drop $\Delta p$.
We consider the interface between the solids at increasing
magnification $\zeta$. At low magnification we observe no surface roughness and it appears
as if the contact is complete. Thus studying the interface only at this low
magnification we would be tempted to conclude that the leak-rate
vanishes. However, as we increase the magnification $\zeta$
we observe surface roughness and non-contact regions, so that
the contact area $A(\zeta)$ is smaller than the nominal contact area $A_0 = A(1)$. As we increase
the magnification further, we observe shorter wavelength roughness, and $A(\zeta)$ decreases
further. For randomly rough surfaces, as a function of increasing magnification, when
$A(\zeta)/A_0 \approx 0.42$ the non-contact area percolate\cite{Dapp}, and the first open channel is observed,
which allow fluid to flow from the high pressure side to the low pressure side.
The percolating channel has a most narrow constriction over which most of the pressure drop
$\Delta p$ occurs. In the simplest picture one assume that the whole
pressure drop $\Delta p$ occurs over this {\it critical constriction}, and if it is
approximated by a rectangular pore of height $u_{\rm c}$ much smaller than its width $w$
(as predicted by contact mechanics theory), the leak rate for a viscous fluid can be 
approximated by\cite{Yang,Boris,LP1,Carbone,Pleak,MusLeak,ComLeak}
$$ \dot Q = {u_{\rm c}^3 \over 12 \eta}\Delta p\eqno(1)$$
where $\eta$ is the fluid viscosity. The height $u_{\rm c}$ of the critical constriction can
be obtained using the Persson contact mechanics theory
(see Ref. \cite{Persson2,aaa,Carbone1,LP1,Prodanov}). The result (1)
is for a seal with a square nominal contact area. For a rectangular contact area with the length
$L_x$ in the fluid flow direction and $L_y$ in the orthogonal direction, there will be an
additional factor $L_y/L_x$ in (1). In a typical case the seal has a circular (radius $r_0$)
cross section (like for rubber O-rings), and in this case $L_y=2 \pi r_0$ and typically $L_y/L_x >> 1$ in which case the
leak-rate will be much larger than given by the {\it square-leak-rate} formula (1). However, this
geometrical correction factor is trivially accounted for. In deriving (1) it is assumed that the fluid
pressure is negligible compared to the nominal contact pressure. If this is not the case one must include the
deformation of the elastic solids by the fluid pressure distribution\cite{liftoff}.

A more general and accurate
derivation of the leakrate is based on the concept of fluid flow conductivity
$\sigma_{\rm eff}$. The fluid flow
current
$$J_x = - \sigma_{\rm eff} {d p_{\rm fluid} \over d x}$$
Since the leakrate $\dot Q = L_y J_x$ we get
$${d p_{\rm fluid} \over d x} = -{\dot Q \over L_y} {1\over \sigma_{\rm eff}}\eqno(2)$$
Note that $\sigma_{\rm eff}$ depends on the contact pressure $p_{\rm con}(x)$ and
hence on $x$. In the present case $p_{\rm fluid} << p_{\rm cont}$ and in this case $p_{\rm cont} \approx p$,
where $p(x)$ is the external applied squeezing pressure (or nominal contact pressure).
Integrating (2) over $x$  gives
$$\Delta p = {\dot Q \over L_y} \int_{-\infty}^{\infty} dx \ {1\over \sigma_{\rm eff}(p(x))}$$
where we have used that $\dot Q$ is independent of $x$ as a result of fluid volume conservation.
For the Hertz contact pressure profile, where $p(x) = 0$ for $x>a$ and $x<-a$,
we get with $y=x/a$:
$$\Delta p = \dot Q {L_x \over L_y} \int_0^1 dy \ {1\over \sigma_{\rm eff}(p (y))}\eqno(3)$$
where $L_x=2a$ is the width of the contact region in the fluid flow direction and where
$$p (y)=p_0 \left (1-y^2\right )^{1/2}\eqno(4)$$
From (3) we get the fluid leakrate (volume per unit time)
$$\dot Q = {L_y \over L_x} {\Delta p \over \int_0^1 dy \ \sigma^{-1}_{\rm eff}(p (y))}\eqno(5)$$

In the calculations presented below we have used 
(5) with the flow conductivity $\sigma_{\rm eff}$ calculated using
the Bruggeman effective medium theory (``corrected'' so the contact area percolate for $A/A_0 = 0.42$; see Ref. \cite{Dapp}),
and the Persson contact mechanics theory (see Ref. \cite{Boris,LP1} for the details).
This theory takes into account all the fluid flow channels and not
just the first percolating channel observed with increasing magnification as in the critical junction theory.
The dependency of the leak rate on the fluid viscosity $\eta$ and
the fluid pressure difference $\Delta p$ given by (1) is the same in the more accurate approach.
Similar, the leak-rate is proportional to $L_y/L_x$ (where $L_x=2a$ in the present case)
in this more accurate approach. 

\vskip 0.5cm
{\bf 3 Leakage of gases}

The leakage of gases is more complex than the leakage of liquids.
The reason is that the separation between the solids in the critical junctions can
be smaller than the mean free path $\lambda$ of the gas atoms (or molecules), due to collision between the
gas atoms, which is of order $\approx 100 \ {\rm nm}$ at room temperature and atmospheric pressures. When $u_{\rm c} < \lambda$
one need to take into account the ballistic motion of the gas atoms between the collisions
with the solid walls. In particular, in experiments where the syringe is surrounded by vacuum
the gas pressure may be much smaller (and the gas atom mean free path much bigger) 
close to the gas exit of the critical junction than that inside the syringe.

When ballistic effects are important one cannot use the fluid continuum equations to study
the flow of the gas through the critical junction. This problem has been studied using the
Boltzmann equation\cite{Bol}. Here we will present two much simpler approaches to derive 
an approximate expression for the leakrate in the ballistic limit. We first briefly
review some well-known facts about the kinetic theory of gases and the continuum 
(hydrodynamic) limit of gas flow.

\vskip 0.2cm
{\bf 3.1 The kinetic theory of gases}

We assume the ideal gas law
$$pV = N k_{\rm B} T,$$
or
$$p = n k_{\rm B} T,$$
where $n=N/V$ is the number of atoms per unit volume.
The average gas atom velocity 
$$\bar v = \left ( {8 k_{\rm B} T\over \pi m}\right )^{1/2} ,\eqno(6)$$
or
$$k_{\rm B} T = {\pi\over 8} m\bar v^2 , \eqno(7)$$
where $m$ is the gas atomic mass.

Let $\sigma$ be the collision cross section of a gas atom. The atom mean free path
$$\lambda = {1\over \surd 2 n \sigma} . \eqno(8)$$
The mean free path $\lambda$ is the distance an atom move (on the average) 
before it makes its first collision with another gas atom. The gas viscosity\cite{kin}
$$\eta = {1\over 3} mn\bar v \lambda . \eqno(9)$$
Using (6) this gives
$$\eta = {2 \over 3} \left ({2\over \pi}\right )^{1/2}  n \lambda \left ( m k_{\rm B} T \right )^{1/2} . \eqno(10)$$
Note that since $\lambda \sim 1/n$ the viscosity is independent of the gas density. 
Qualitatively it can be understood by considering the shearing of two parallel planes 
containing a gas inbetween. If the density of the gas is doubled, there are twice many molecules 
available to transport momentum from one plate to the other, but the mean free path of each molecule is 
also halved, so that it can transport this momentum only half as effectively.

In what follows we will assume that the temperature is constant. At room temperature
for He the mean free path $\lambda \approx 174 \ {\rm nm}$.
Table \ref{tab:gas} gives the viscosity and mean free path for several gases.

\begin{table}[hbt]
   \caption{Viscosity $\eta$, the gas atom mean free path $\lambda$ and the average gas atom (or molecule) 
speed $\bar v$ at room temperature and 1 atm pressure.}
   \label{tab:gas}
   \begin{center}
      \begin{tabular}{@{}|l||c|c|c|@{}}
         \hline
            gas   &  $\eta$ ($10^{-5} \ {\rm Pa s}$) & $\lambda$ (nm) & $\bar v$ \ (m/s) \\
         \hline
         \hline
            He & 1.96 & 174 & 1245 \\
         \hline
            N$_2$ & 1.76 & 59 & 470\\
         \hline
            O$_2$  & 2.04 & 63 & 440 \\
         \hline
            CO$_2$  & 1.47 & 39 & 375 \\
         \hline
      \end{tabular}
   \end{center}
\end{table}

\begin{figure}
        \includegraphics[width=0.4\textwidth]{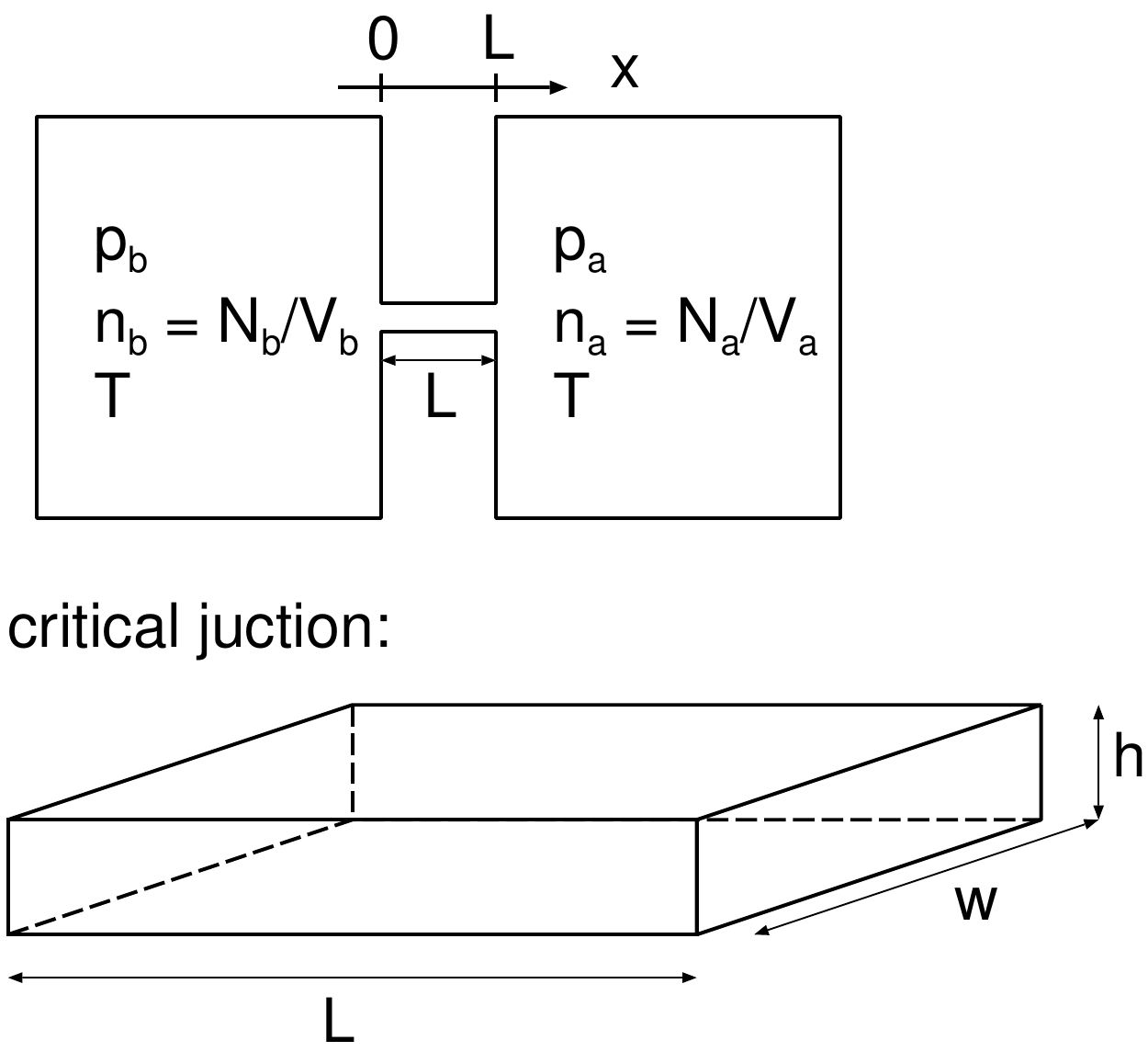}
        \caption{\label{CriticalJunctionPic.eps}
A rectangular constriction between two volumes with gas at pressures
$p_{\rm a}$ and $p_{\rm b}$.
}
\end{figure}

\begin{figure}
        \includegraphics[width=0.45\textwidth]{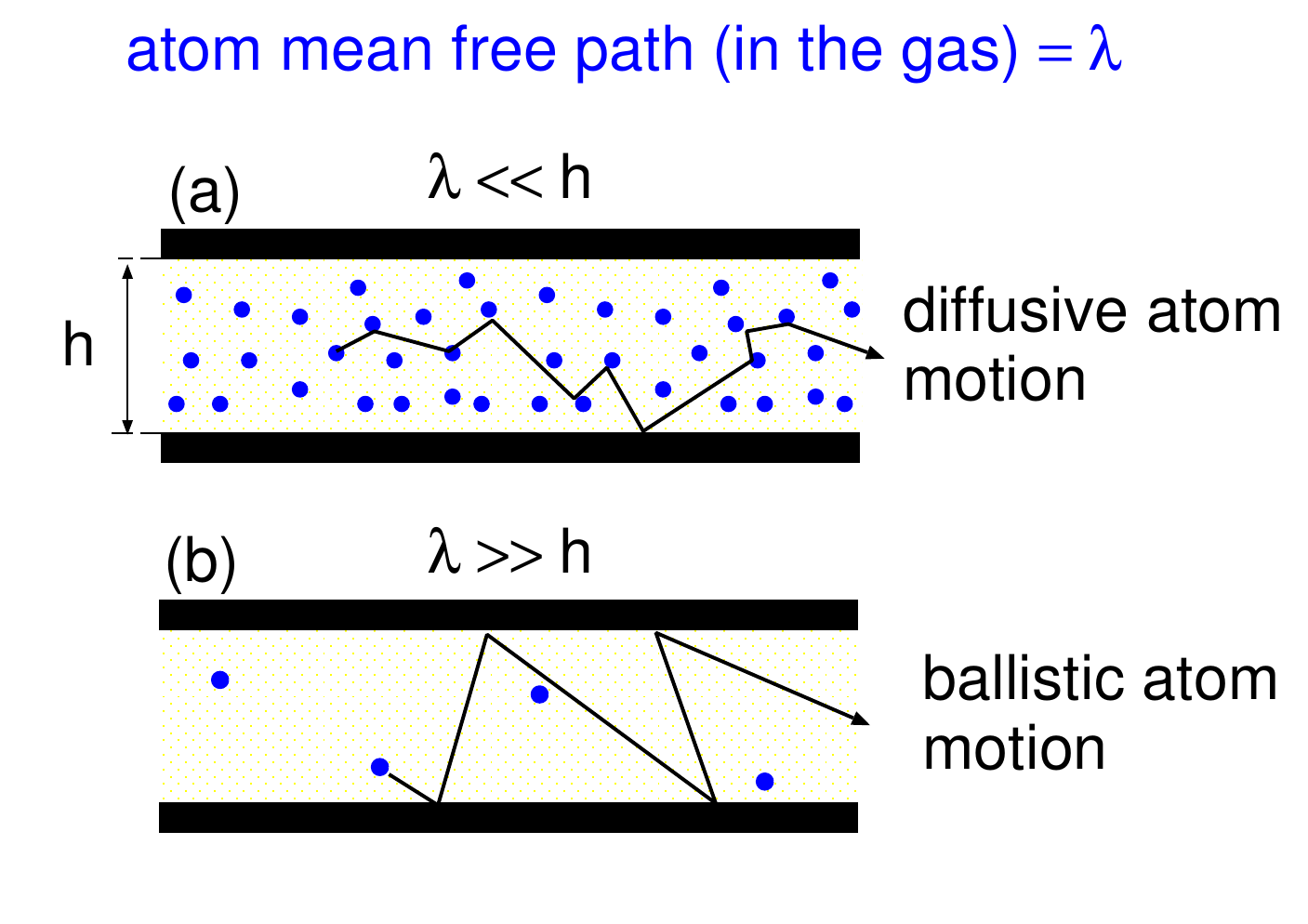}
        \caption{\label{Ballistic.eps}
Diffusive (a) and ballistic (b) motion of the gas atoms (e.g. He) in the critical junction.
In case (a) the gas mean free path $\lambda$ is much smaller than the gap width $h=u_{\rm c}$ and the gas
molecules makes many collisions with other gas molecules before a collision with the solid walls.
In the opposite limit, when $\lambda >> u_{\rm c}$ the gas molecules makes many collisions with the solid wall before colliding
with another gas molecule. In the first case (a) the gas can be treated as a
(compressible) fluid, while a kinetic approach is needed in case (b).}
\end{figure}

\vskip 0.3cm
{\bf 3.2 Fluid flow through rectangular constriction}

Let us assume that the 
height $h$ of the rectangular constriction is much smaller than
its width $w$ and length $L$ (see Fig. \ref{CriticalJunctionPic.eps}). 
Let $x$ be a coordinate axis along the constriction and assume the gas pressure 
$p=p_{\rm b}$ for $x=0$ and $p=p_{\rm a}$ for $x=L$. The atom number densities on the two sides are
denoted $n_{\rm b}$ and $n_{\rm a}$.

We consider two limits, namely $\lambda << h$ and $\lambda >> h$, where $\lambda$ is the mean free path. 
The first limit $\lambda << h$ can be studied using continuum fluid flow dynamics for a compressible fluid.
In the other limit $\lambda >> h$ ballistic flow occurs, where one can neglect the collisions 
between gas atoms and assume the gas atoms collide only with the solid walls (see Fig. \ref{Ballistic.eps}).
Note that ballistic refer only to the absence of collisions between the air molecules; the
molecules may still scatter diffusely from the solid walls so the motion of gas molecules through the junction is
always diffusive-like.

\vskip 0.2cm
{\bf Continuum flow limit $\lambda << h$}

For fluid flow between two parallel flat surfaces, if the separation $h << w$ and $h << L$,
then the fluid velocity will vary much more rapidly in the normal direction $z$ then in the fluid
flow direction $x$. In this case the Navier-Stokes equation reduces to
$$ {\partial \over \partial z} \left ( \eta {\partial v_x \over \partial z} \right ) \approx {\partial p \over \partial x},\eqno(11)$$
and the fluid continuity equation
$${\partial \over \partial x} (nv_x) = 0,\eqno(12)$$
where $n$ is the gas number density (so $\rho = m n$ the gas mass density).
Assuming an ideal gas the viscosity $\eta$ is independent of the gas density so that (11) takes the form
$$ \eta {\partial^2 v_x \over \partial z^2} \approx {\partial p \over \partial x}.\eqno(13)$$
Since for an ideal gas $p=n k_{\rm B} T$ we can write (12) as 
$${\partial \over \partial x} (p v_x) = 0,$$
where we have assumed that the temperature is constant. We expect $p$ to depend mainly on $x$ and
from (13) we get
$$v_x \approx {1\over 2 \eta} z^2 {\partial p \over \partial x} +Az+B$$
Since $v_x = 0$ for $z=0$ and $z=h$ we get
$$v_x \approx {1\over 2 \eta} z (z-h) {\partial p \over \partial x}$$
The flow current 
$$J_x = \int_0^h d z \ n v_x \approx - {h^3\over 12 \eta} n {\partial p \over \partial x}   $$
must be independent of $x$, and using
$p=n k_{\rm B} T$ this gives
$$p {\partial p \over \partial x} = C$$
or
$$p^2 = 2 C x +D$$
Assuming $p=p_{\rm b}$ for $x=0$ and $p=p_{\rm a}$ for $x=L$ we get
$$p^2 = (p_{\rm a}^2-p_{\rm b}^2) (x/L)+p_{\rm b}^2$$
Hence
$$J_x \approx - {h^3 \over 12 \eta} {p\over k_{\rm B} T} {\partial p \over \partial x} 
= {h^3 \over 24 \eta L} {p_{\rm b}^2-p_{\rm a}^2\over k_{\rm B} T}$$
The number of atoms moving through the constriction per unit time equal
$\dot N = wJ_x$ or
$$\dot N = {h^3 w \over 24 \eta L} {p_{\rm b}^2 - p_{\rm a}^2\over k_{\rm B}T}\eqno(14)$$

In some leakage tests the syringe filled with gas (e.g. He) is put in vacuum. In this case
$p_{\rm a} = 0$ so that
$$\dot N = {h^3 w \over 24 \eta L} {p_{\rm b}^2 \over k_{\rm B}T}$$
Using $p_{\rm b} = n_{\rm b} k_{\rm B} T$ and (6) and (9) this gives
$$\dot N = {\pi \over 64} n_{\rm b} \bar v{h^3 \over \lambda_{\rm b}} {w\over L} \eqno(15)$$

\begin{figure}
        \includegraphics[width=0.2\textwidth]{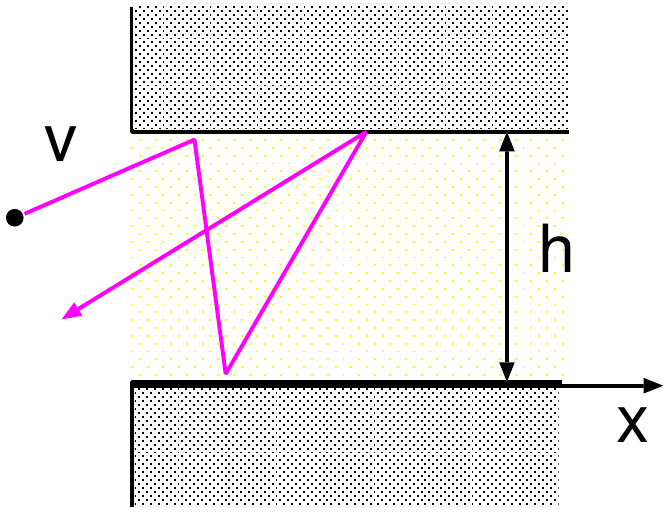}
        \caption{\label{BackReflacted.eps}
A gas atom which enter a constriction can be back scattered due to the atomic corrugation of the solid walls, or 
due to surface roughness or the thermal motion of the wall atoms.
}
\end{figure}

\begin{figure}
        \includegraphics[width=0.45\textwidth]{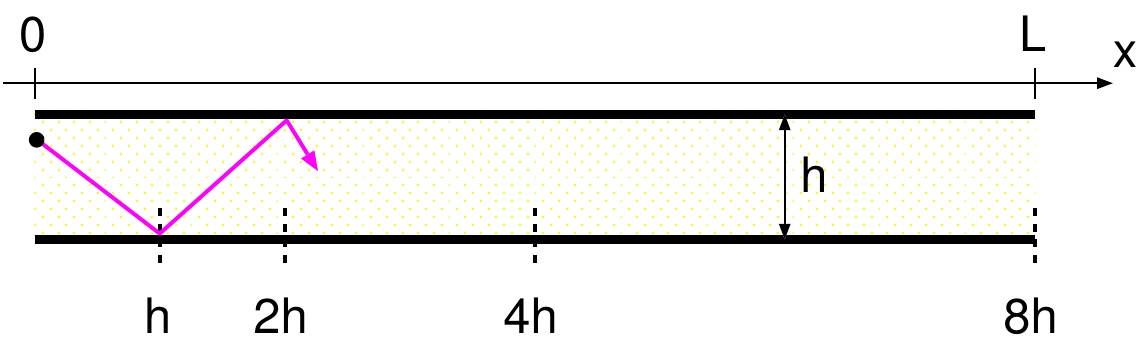}
        \caption{\label{Moving.eps}
An atom in the constriction collide with the walls in the channel at locations separated 
in the $x$-direction with on the average the distance $\approx h$. We assume diffusive scattering of the
atom from the walls so when an atom is at $x=h$ (where it experience its first collision with a wall)
it has equal probability to return to $x=0$ as to move to
$x=2h$. Hence the probability that it moves to $x=2h$ is $1/2$. When it is at $x=2h$ it has equal probability
to return to $x=0$ as to move to $x=4h$. Hence the probability to move to $x=4h = 2^2 h$ will be $(1/2)\times (1/2) = (1/2)^2$.
In the same way, the probability to move to $x=2^n h$ will be $(1/2)^n$. If it arrives to $x=L$ it is transferred
from the high pressure side to the low pressure side. Thus the probability that an atom which enter the
constriction at $x=0$ will exit it at $x=L$ will be $P=(1/2)^n$ where $2^n h=L$. Hence $P\approx h/L$. 
}
\end{figure}

\vskip 0.2cm
{\bf Ballistic flow: qualitatively argument}

We consider the flow of gas atoms through the critical junction in the ballistic limit $\lambda >> h$.
Here we present a simple hand-waving argument: 
The flow of atoms into the junction from the high pressure side is $(1/4) n_{\rm b} \bar v$ times the area
of the constriction given by $wh$. 
(The factor of 1/4 is the result of the fact that only half of the atoms moves in the positive $x$-direction, and due to the fact
that on the average they only move in the $x$-direction with the speed $\bar v/2$.)
Thus the number of atoms flowing into the constriction per unit time is $(1/4) n_{\rm b} \bar vwh$. However,
the gas atoms will scatter from the walls in a random way (diffusive scattering) due to atomic surface roughness on the walls, and also due to the
random thermal motion of the wall atoms. Thus some of the atoms which enter will get back-scattered and return to the high 
pressure volume $V_{\rm b}$ (see Fig. \ref{BackReflacted.eps}). 
It turns out only a fraction $2 h/L$ of the atoms will be able to penetrate the junction (see Fig. \ref{Moving.eps}). 
Thus the number of atoms per unit time moving through the constriction will be
$$\dot N = {1\over 2} n_{\rm b} \bar v h^2 {w\over L}\eqno(16)$$ 

\vskip 0.2cm
{\bf Ballistic flow: diffusion equation}

We now present a more accurate derivation of the result (16).
When $\lambda >> h$ the leakage through a critical junction can be determined by a simple kinetic approach.
We consider a junction with the height $h$, the width $w$ and the length (in the leakage flow direction) $L$.
The average collision time for an atom in the gap with the walls is $\tau$, 
and we expect $\bar v \tau \approx h$, where $\bar v$ is the average
atom velocity. Due to surface roughness, when an atom hit the wall it is reflected in a nearly random direction 
so sometimes in the leakage direction and sometimes in the opposite direction. The atom moves on the average a distance 
$\Delta x\approx h$ along or opposite to the leakage direction between each collisions with the wall.  
Let us use a discretized (in time and space) model and 
consider the probability $P(x,t)$ to find the particle at $x$ at time $t$. If the particle is at $x$ at time $t+\tau$ the particle
must have been at either $x-\Delta x$ (with probability $1/2$) or at $x+\Delta x$ (with probability $1/2$) at time $t$. Thus we get
$$P(x,t+\tau) = {1\over 2} P(x-\Delta x,t)+ {1\over 2} P(x+\Delta x,t)$$
If we expand the probability to linear order in $\tau$ and quadratic order in $\Delta x$ we get a diffusion equation for $P(x,t)$:
$${\partial P \over \partial t} = {\Delta x^2\over 2 \tau} {\partial^2 P\over \partial x^2}$$
Since the gas atoms does not interact with each other in the junction, we can also consider $P(x,t)$ as the
concentration of atoms in the gap. If we denote this by $c(x,t)$ and if we use $\Delta x\approx h$
and $\bar v \tau \approx h$ we get
$${\partial c \over \partial t} - D {\partial^2 c\over \partial x^2}=0\eqno(17)$$
where the diffusivity $D= \bar v u_{\rm c}/2$. The gas flow current $J$ satisfies the continuity equation
$${\partial c \over \partial t} + {\partial J \over \partial x}=0\eqno(18)$$
Comparing (17) and (18) gives
$$J=-D{\partial c\over \partial x}\eqno(19)$$
Here we are interested in a stationary state so that $c(x,t)$ is time independent. From (17) we get
$${\partial^2 c\over \partial x^2}=0\eqno(20)$$
i.e., $c(x)$ is a linear function of $x$.
Assuming $c=n_{\rm b}$ for $x=0$, where $n_{\rm b}$ is the number of atoms per unit volume of the gas
on the inlet side, and assuming that $c$ vanish (vacuum) on the exit side $x=L$, we get from (20):
$$c(x)=n_{\rm b} \left (1-{x\over L}\right ).$$
Using (19) this gives the current $J=Dn_{\rm b}/L$. The gas leakage $\dot N$ 
is determined by the product between the current $J$ and the
junction cross section area $w h$ so that $\dot N = J w h = D n_{\rm b} h w/L$.
Using $D=\bar v h/2$ this gives
$$\dot N = {1\over 2} n_{\rm b} \bar v h^2 {w\over L}, \eqno(21)$$
which agree with (16).

In the derivation of (21) we have assumed a rectangular junction and that the
molecules scatter randomly from the solid walls. Real junctions are not perfectly
rectangular and the gas atoms may not scatter at random 
directions from the solid walls. Still, the theory we use gives good agreement with
experiments which indicate that the idealized treatment is close to reality.

\vskip 0.2cm
{\bf Interpolation formula and comparison with experiments}

We can interpolate between the limits (15) and (21) using
$$\dot N = {\pi \over 64} {w\over L} h^3 n_{\rm b} \bar v \left ( {1\over \lambda_{\rm b}} + {\xi \over h}\right )\eqno(22)$$ 
where $\xi = 32/\pi$. 
We can interpret the factor
$${1\over \lambda_{\rm eff}} = {1\over \lambda_{\rm b}} + {\xi \over h}\eqno(23)$$
as defining an effective mean free path. 

The formula (22) gives leakage rates in close agreement with 
experiments and with the Boltzmann theory predictions assuming diffusive atom-wall scattering.
Thus in Ref. \cite{Nacer0,Nacer1} experimental results was presented for rectangular channels with $w/h \approx 3$
but $L>> w$. Our calculation is for $w/h >> 1$ but theories have been developed for finite $w/h$
ratio\cite{Nacer0}, and taking this into account 
the experimental results presented in Ref. \cite{Nacer1} is in good agreement with our theory prediction.

\vskip 0.2cm
{\bf Application to syringe He-gas leakage testing}

Helium is a very inert gas and it will not react with materials 
but it may interdiffuse in some materials like polymers or rubber compounds. 
In most helium leak testing applications, one uses a mass spectrometer tuned to detect helium.
High vacuum testing allows leak test down as low as 
$\dot V \approx 10^{-12} \ {\rm cm}^3/{\rm s}$ of He of atmospheric pressure.

Suppose we put a syringe (with closed needle) filled with He gas (volume $V_{\rm b}$ at pressure $p_{\rm b}$) in a closed volume $V_{\rm a}$
at pressure $p_{\rm a}=0$ (vacuum). Due to He gas leakage the pressure outside of the syringe will slowly increase.
Since $pV=N k_{\rm B} T$ we get $\dot p_{\rm b} V_{\rm b} = \dot N_{\rm b} k_{\rm B} T$ 
and $\dot p_{\rm a} V_{\rm a} = \dot N_{\rm a} k_{\rm B} T$,
where we have assumed that the temperature is constant.
Since the gas atoms leaving the syringe enters the volume $V_{\rm a}$ we must have (conservation of atoms) $\dot N_{\rm a}=-\dot N_{\rm b}$
so that: 
$$\dot  p_{\rm a} V_{\rm a} = - \dot p_{\rm b} V_{\rm b} =  \dot N_{\rm a} k_{\rm B} T\eqno(24)$$
To apply this to syringes we use (22) with $w=L$ (square shaped critical junction), $h=u_{\rm c}$ and with an
additional factor $L_y/L_x \approx 100$ due to the rectangular nature of the nominal rib-glass contact region.
Thus the gas atom leakrate
$$\dot N \approx {\pi \over 64} u_{\rm c}^3 n_{\rm b} \bar v {L_{\rm y} \over L_{\rm x}}
\left ( {1\over \lambda_{\rm b}} + {\xi \over u_{\rm c}}\right )\eqno(25)$$ 
and (24) takes the form
$$\dot  p_{\rm a} V_{\rm a} \approx {\pi \over 64} u_{\rm c}^3 n_{\rm b} \bar v 
{L_{\rm y} \over L_{\rm x}} \left ( {1\over \lambda_{\rm b}} + {\xi \over u_{\rm c}}\right )k_{\rm B} T\eqno(26)$$
Using that $p_{\rm b} = n_{\rm b} k_{\rm B}T$ we can also write (26) as
$$\dot  p_{\rm a} V_{\rm a} \approx {\pi \over 64} u_{\rm c}^3 p_{\rm b} \bar v 
{L_{\rm y} \over L_{\rm x}} \left ( {1\over \lambda_{\rm b}} + {\xi \over u_{\rm c}}\right ) \eqno(27)$$
In many syringe applications $\xi / u_{\rm c} >> 1/\lambda_{\rm b}$ and in this case (27) reduces to
$$\dot  p_{\rm a} V_{\rm a} \approx {1 \over 2} u_{\rm c}^2 p_{\rm b} \bar v 
{L_{\rm y} \over L_{\rm x}} \eqno(28)$$

The quantity $\dot p_{\rm a} V_{\rm a} = \dot N_{\rm a} k_{\rm B}T$, 
with units ${\rm Pa \times m^3 /s}$ or ${\rm mbar \times liter / s}$, is often measured in He leakage experiments\cite{Sarah1,Sarah2,Alej}. 
Sometimes the leakrate is given as volume of gas at the pressure $p_{\rm b}$
passing through the constriction per unit time. This is obtained from $\dot N$ using $V = N k_{\rm B}T/p_{\rm b}$ so that
$\dot V = \dot N k_{\rm B}T/p_{\rm b}$, where we have assumed that $p_{\rm b}$ is constant. Often $\dot V$ is given as ${\rm cm^3/s}$.
Using (25) and $p_{\rm b} =n_{\rm b} k_{\rm B}T$ we get
$$\dot V \approx {\pi \over 64} u_{\rm c}^3  \bar v {L_{\rm y} \over L_{\rm x}}
\left ( {1\over \lambda_{\rm b}} + {\xi \over u_{\rm c}}\right )\eqno(29)$$ 
If $\xi / u_{\rm c} >> 1/\lambda_{\rm b}$ (29) reduces to
$$\dot V \approx {1 \over 2} u_{\rm c}^2  \bar v {L_{\rm y} \over L_{\rm x}}\eqno(30)$$ 

We have observed (see also Ref. \cite{Sarah2}) that syringes gives typical 
leakrates $\dot p_{\rm a} V_{\rm a} = (1-100) \times 10^{-8}  \ {\rm mbar \times liter /s}$. 
If we assume $p_{\rm b} = 0.1 \ {\rm MPa}$ (atmospheric pressure) and $p_{\rm a} \approx 0$ (vacuum) 
and $L_y/L_x = 100$ we get from (28) that $u_{\rm c} \approx 4 \ {\rm nm} $ 
if $\dot p_{\rm a} V_{\rm a} = 10^{-6} \ {\rm mbar \times liter /s}$,
$u_{\rm c} \approx 1 \ {\rm nm}  $ if $\dot p_{\rm a} V_{\rm a} = 10^{-7} \ {\rm mbar \times liter /s}$.

He atoms (and air molecules) can be absorbed by the rubber matrix and 
diffuse through the rubber stopper, and this too will contribute to the observed leakage. 
Let us estimate the leakage rate due to He-diffusion through the
rubber. We consider a rib on the stopper which we treat as a rectangular region
of height $d_0$ and length (in the gas diffusion direction which we take as the $x$-direction) 
$d_1$ and width $L_y = 2 \pi R$. Hence the cross section area $A=d_0 L_y$. 
We assume a steady state (see below) where the 
concentration of He atoms varies linearly
between the inner surface of the rubber stopper and the outer surface. 
The He diffusion current
$$J=-D_{\rm g} {d c \over dx} = D_{\rm g} {c_{\rm b}-c_{\rm a}\over d_1}$$
where the gas diffusion coefficient $D_{\rm g}$ has the unit ${\rm m^2/s}$.
We assume that the concentration of He atoms at the inner and outer surfaces are proportional to the
gas pressures inside and outside the syringe. In this case $c_{\rm b} = S p_{\rm b}$ 
(where $S$ is the solubility) and $c_{\rm a} = 0$
since we assume $p_{\rm a} = 0$ (vacuum). Thus we get
$$\dot N = J A = D_{\rm g}S {A\over d_1} p_{\rm b}$$
The quantity $\kappa=D_{\rm g}S k_{\rm B}T$ (with units ${\rm m^2/s}$) is denoted the gas permeability. 
Using $\dot p_{\rm a} V_{\rm a} = \dot N k_{\rm B} T$ we get
$$\dot p_{\rm a} V_{\rm a} = \kappa {A\over d_1} p_{\rm b}$$ 


Depending on the type of rubber, for He 
$\kappa =(3-30)\times 10^{-12} \ {\rm m^2 /s} $ (see Ref. \cite{Diffusivity}).
while for ${\rm N}_2$ the gas permeability may be $\sim 10$ times smaller.
We consider the He gas leakage through a rib treated as a rectangular region with height $d_0$,
length (in the leakage direction) $d_1$ and width $L_y=2 \pi R$ we get $A/d = 2 \pi R d_0/d_1$.
Typically, $d_0\approx d_1$ so we get
$$\dot p_{\rm a} V_{\rm a} \approx \kappa 2 \pi R p_{\rm b}\eqno(32)$$
Using $R=0.5 \ {\rm cm}$  this gives
$$\dot p_{\rm a} V_{\rm a} \approx 10^{-7} - 10^{-6}  \ {\rm mbar \cdot liter /s}$$
This is similar to the measured leakage rates for syringes. Thus, in many cases the leakage
will be strongly influenced by diffusion of He through the rubber matrix.

Finally, let us determine how long time it takes for diffusion through the rubber 
to reach the steady state. Assume that the
He gas in introduced in the syringe at time $t=0$. From the diffusion equation we know 
(from dimensional arguments) that it will take the time $t_0 \approx d_1^2/D_{\rm g}$ to reach the steady state.
For He in rubber $D_{\rm g} \approx 10^{-10} \ {\rm m^2/s}$ and with $d_1 \approx 1 \ {\rm mm}$ this gives
$t_0 \approx 10^4 \ {\rm s}$. A more accurate analysis gives a factor $\sim 6$ shorter time so we conclude
it typically take $\approx 1 \ {\rm hour}$ to reach the steady state. In most leakage studies the observation time might
be much smaller than this, and in these cases a smaller (and time dependent) leakrate will be observed.
We note for $t < 0.1 d_1^2/D_{\rm g}$ theory predict negligible leakage resulting from He diffusion through the rubber. 
On the other hand the gas leakage through the open (non-contact) channels at the rubber-barrel
interface occurs immediately after filling the syringe with gas. It is clear that the time-dependency of the
leakrate may contain very important information about the origin of the leakage.

We note that leakage of air through most types of rubber is slower than for He.
One well studied case is the diffusion of different gas molecules in ethylene-propylene-diene (EPDM) elastomer\cite{EPDM}.
The diffusivity at room temperature for He and ${\rm N}_{\rm 2}$ in EPDM are 
$D_{\rm g} = 1.7 \times 10^{-9}$ and $5.5 \times 10^{-11} \ {\rm m^2/s}$. The higher diffusivity for He reflects its smaller effective size
(effective van der Waals diameter of He and ${\rm N}_{\rm 2}$ are $2.8 \ {\rm nm}$ and $3.8  \ {\rm nm}$, respectively). However,
the solubility for He in EPDM is $\approx 5$ times lower than for ${\rm N}_{\rm 2}$ so the He permeability $\kappa=D_{\rm g}S k_{\rm B}T$
is only $\sim 6$ times bigger than for ${\rm N}_{\rm 2}$.

\begin{figure}
        \includegraphics[width=0.49\textwidth]{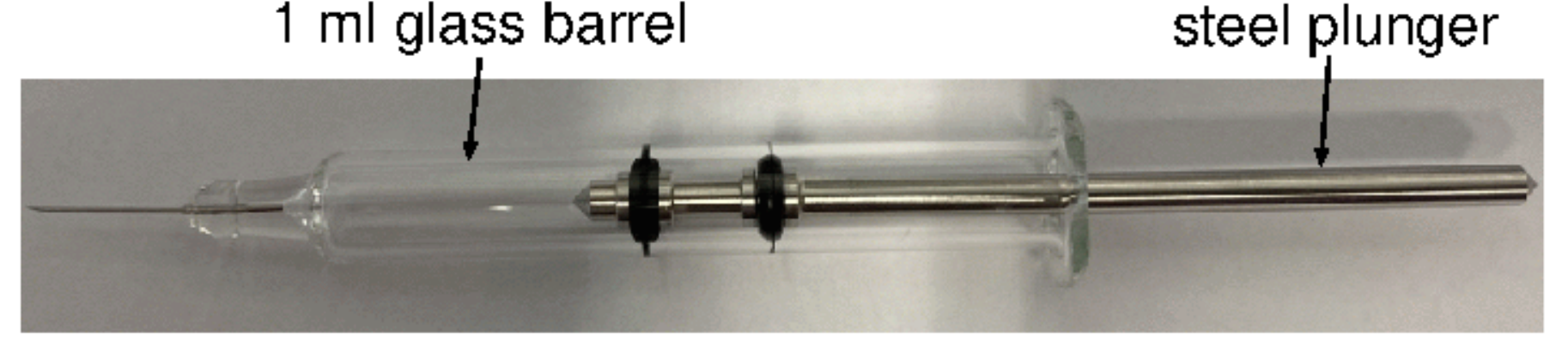}
        \caption{\label{NewAlStopper.eps}
Steel plunger with two rubber NBR O-rings in a glass barrel.
}
\end{figure}

\vskip 0.2cm
{\bf 4 Leakage experiments with syringes with rubber O-ring stopper}

Most syringes exhibit gas leakage which is too small to be accurately studied and which
may result from diffusion of gas through the rubber, or of some other origin.
To obtain larger leakage rates experiments have been performed where a thin metal wire
is located between the rubber stopper and the barrel\cite{wire}. Here we instead
use rubber stoppers which are sandlasted to produce random roughness. 
We used a steel plunger with two rubber O-rings and a 1 ml glass barrel (see Fig. \ref{NewAlStopper.eps}). 
We use O-rings with the inner diameter $D=3 \ {\rm mm}$, and the rubber cross section diameter $2R=2 \ {\rm mm}$.
Hence the O-ring outer diameter $d=D+4R=7  \ {\rm mm}$. However, the sandblasted O-rings have smaller outer diameter.
We have sandblasted O-rings made from 6 different types of
rubber (exhibiting very different wear rates\cite{sand}). 
We have measured the surface topography of the sandblasted
surfaces. We have also measured the viscoelastic modulus of
the O-rings rubber compounds. Here we show the results obtained 
for O-rings made from nitrile butadiene rubber (NBR70).

\begin{figure}
        \includegraphics[width=0.45\textwidth]{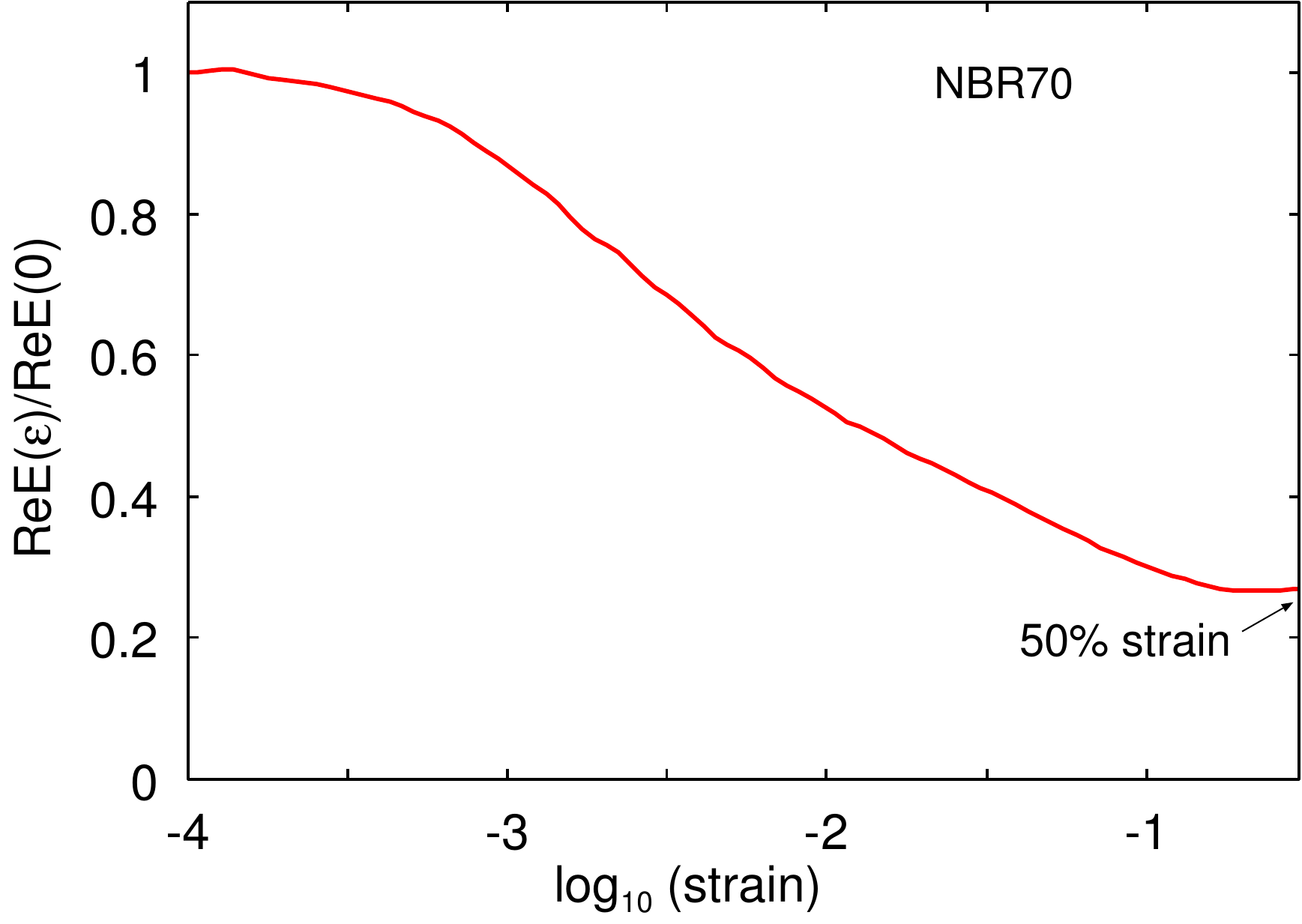}
        \caption{\label{1logStrain.2RatioE.red.NBR70.green.EPDM70.eps}
Strain softening. The ratio between the real part of the viscoelastic modulus
for zero strain and the strain $\epsilon$, as a function of the logarithm of the strain.
	The results is for NBR70 at $T=20^\circ {\rm C}$ and for the frequency $f=1 \ {\rm Hz}$.
}
\end{figure}

\vskip 0.2cm
{\bf 4.1 Rubber viscoelastic modulus}

We use rubber O-rings made from nitrile butadiene rubber (NBR70) with the glass transition temperature $T= -34.3 ^\circ {\rm C}$.
The low strain and low frequency ($\omega \approx 10^{-3} \ {\rm s}^{-1}$,
corresponding to the waiting time $t \approx 1 \ {\rm hour}$) elastic modulus 
is about $25 \ {\rm MPa}$ but due to strain softening for a more relevant strain (of order $50\%$)
the effective modulus is $\approx  7 \ {\rm MPa}$ (see Fig. \ref{1logStrain.2RatioE.red.NBR70.green.EPDM70.eps}). 
As will be shown below, in the analysis presented the actual value of the elastic modulus is irrelevant as long as the strain
in the (macroscopic) nominal contact area is very similar to the typical strain involved in the asperity contact regions.

\begin{figure}
        \includegraphics[width=0.45\textwidth]{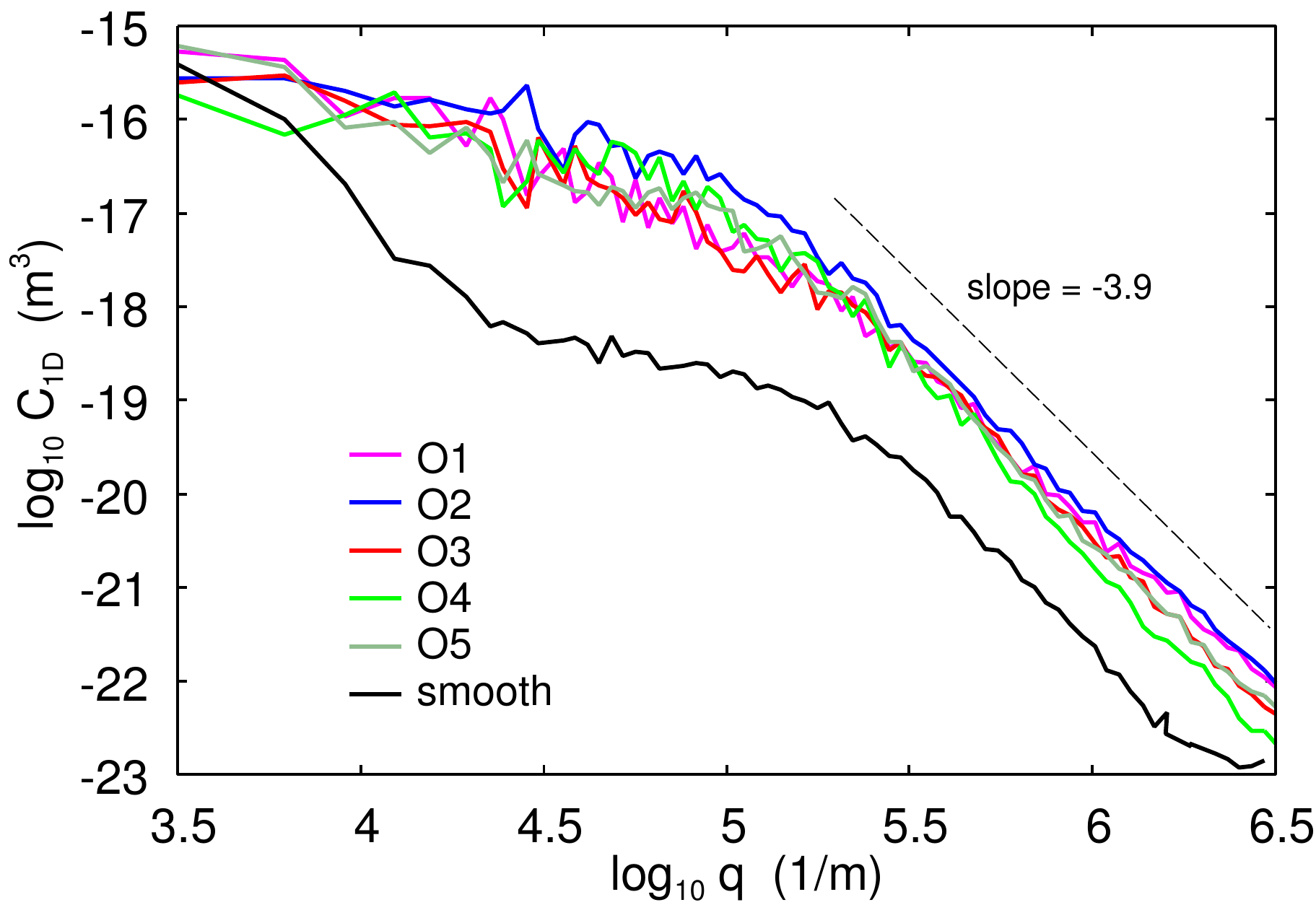}
        \caption{\label{1logq.2logC1D.O1O2O3.good.eps}
	The 1D surface roughness power spectra of the sandblasted O-rings O1-O5,
	and for a not sandblasted (smooth) O-ring.
	}
\end{figure}

\begin{figure}
        \includegraphics[width=0.45\textwidth]{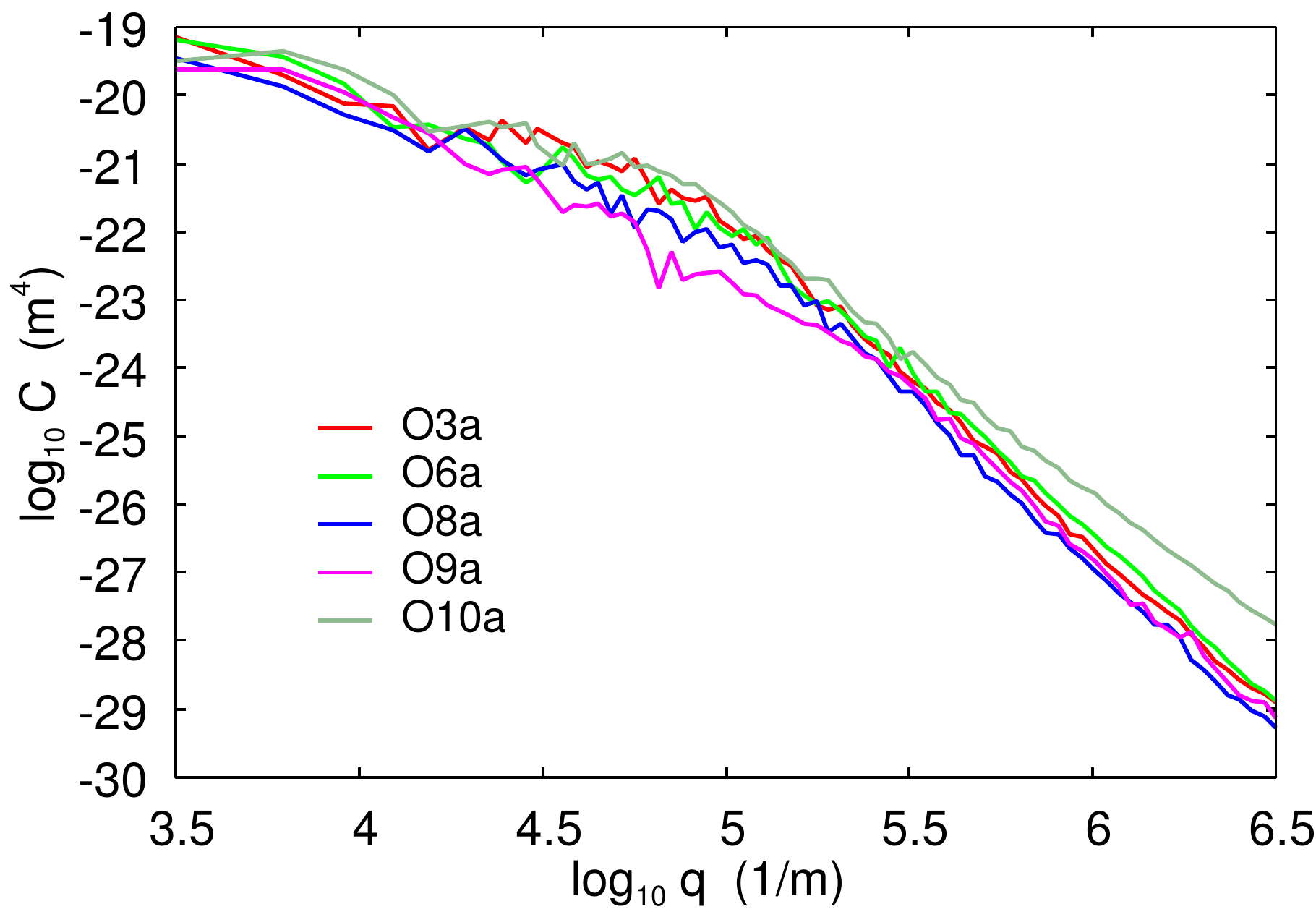}
        \caption{\label{1logq.2logC2D.O3a.O6a.O8a.O9a.O10a.eps}
       The 2D surface roughness power spectra obtained from 1D line scans from the sandblasted rubber O-rings
	O3a, O6a, O8a, O9a, O10a. In each case the power spectra was averaged over 7 line scans each $2 \ {\rm mm}$ 
	long in the direction of the largest O-ring curvature radius. 
	}
\end{figure}

\begin{figure}
        \includegraphics[width=0.45\textwidth]{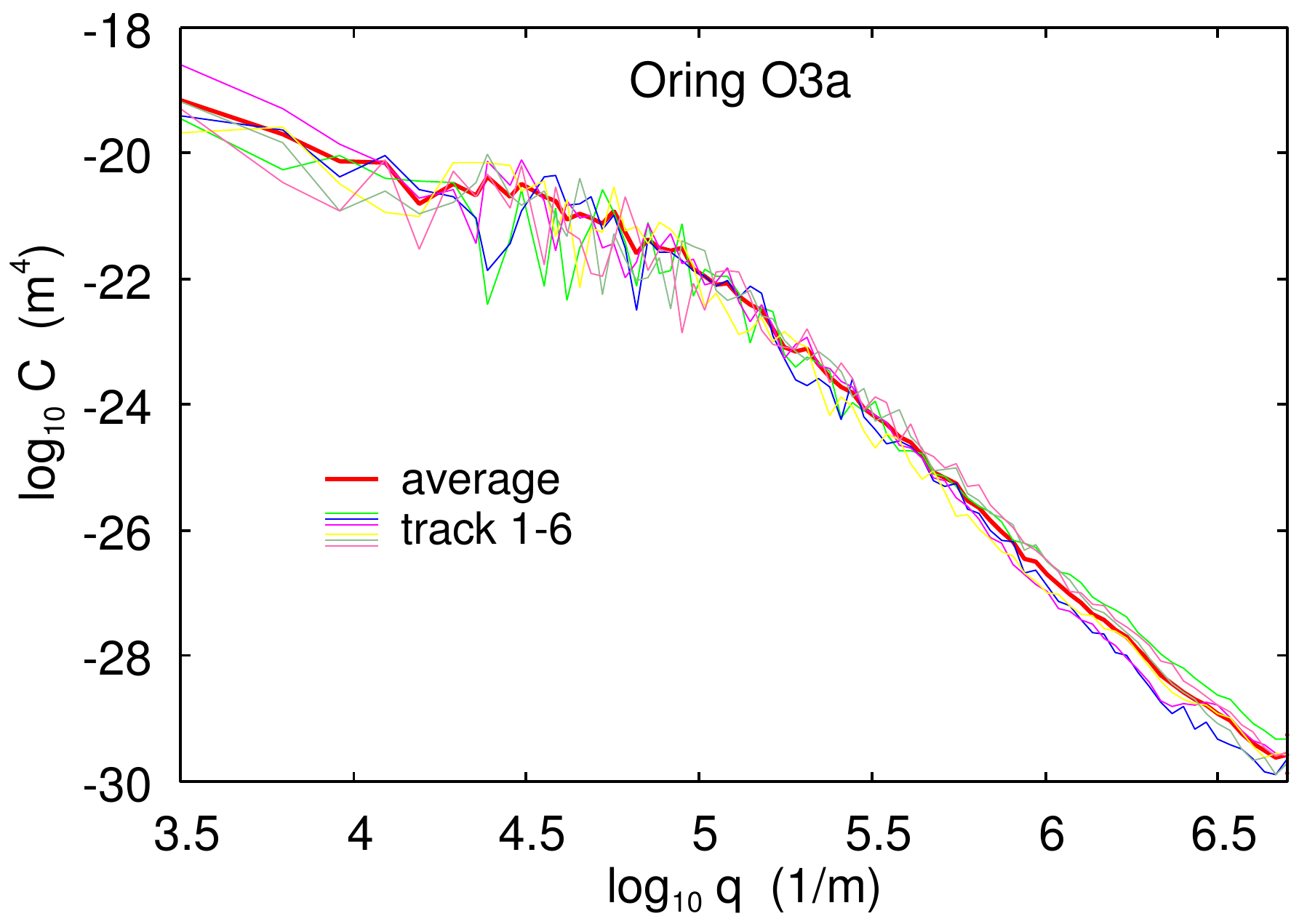}
        \caption{\label{1logq.2logC2D.O3a.all1.eps}
       The 2D surface roughness power spectra obtained from 1D line scans from the sandblasted rubber O-ring
	O3a. The thick red line is the power spectrum obtained by averaging the power spectra obtained
	from 7 line scans each $2 \ {\rm mm}$ 
	long in the direction of the largest O-ring curvature radius. Then thin lines gives the power spectra of each
	individual line scan.
	}
\end{figure}

\begin{table}[hbt]
	\caption{
	The root-mean-square (rms) roughness amplitude $h_{\rm rms}$ (in ${\rm \mu m}$) and the rms 2D-slope for the studied O-rings. 
}
   \label{tab:oringO}
   \begin{center}
      \begin{tabular}{@{}|l||c|c|@{}}
         \hline
	      O-ring   &  $ h_{\rm rms}$  (${\rm \mu m}$) & rms-slope  \\
         \hline
         \hline
	      smooth & 2.21 & 0.09  \\
         \hline
	      O1 & 3.69 & 0.36  \\
         \hline
	      O2 & 4.46 & 0.48  \\
         \hline
	      O3 & 3.02 & 0.33  \\
         \hline
	      O4 & 3.46 & 0.34  \\
         \hline
	      O5 & 3.50 & 0.33  \\
         \hline
	      O3a &  5.80 & 0.51  \\
         \hline
	      O6a &   5.02 & 0.48 \\
         \hline
	      O8a  &   3.81 & 0.36  \\
         \hline
	      O9a &   3.45 & 0.30  \\
         \hline
	      O10a &   6.34 & 0.87 \\
         \hline
      \end{tabular}
   \end{center}
\end{table}

\vskip 0.2cm
{\bf 4.2 Sandblasting and rubber surface roughness}

We had sandblasted two sets of 10 NBR70 rubber O-rings which we denote by
O1-O10 and O1a-O10a (the two sets of O-rings was sandblasted at different days). 
The sandblasting was performed by putting a rubber
O-ring on a steel rod which is attached to an electric drill which rotates the steel rod at about 1 rotation per second,
while it was exposed to a beam of spherical glass particles with a diameter of order a few micrometer. The beam of particles was 
centered on the rubber O-ring and the sandblasting air pressure was about $10 \ {\rm bar}$, and the sandblasting time was varied
between 10 \ {\rm minutes} and  30 \ {\rm minutes} in order to obtain O-rings with different surface roughness and effective radius.

We have measured the surface roughness profile using a stylus instrument 
[Mitutoyo Portable Surface Roughness Measurement Surftest SJ-410 with a
diamond tip with the radius of curvature $r_0 =1 \ {\rm \mu m}$, and with the tip-substrate repulsive force $F_{\rm N} = 0.75 \ {\rm mN}$
and the tip speed $v=50 \ {\rm \mu m/s}$],
and calculated the surface roughness
power spectrum as described in detail elsewhere\cite{Ref6}.
Fig. \ref{1logq.2logC1D.O1O2O3.good.eps} shows the 1D surface roughness power spectra of the sandblasted O-rings O1-O5,
and for a not sandblasted O-ring. 
Fig. \ref{1logq.2logC2D.O3a.O6a.O8a.O9a.O10a.eps}
shows the 2D surface roughness power spectra obtained from 1D line scans from the sandblasted rubber O-rings
O3a, O6a, O8a, O9a, O10a. 
In each case the power spectra was obtained by averaging over 7 line scans, each $2 \ {\rm mm}$ 
long in the direction of the largest O-ring curvature radius. 

Fig. \ref{1logq.2logC2D.O3a.all1.eps}
shows the 2D surface roughness power spectra obtained from 1D line scans from the sandblasted rubber O-ring
O3a. The thick red line is the average power spectrum, obtained by average the power spectra obtained
from 7 line scans. The thin lines gives the power spectra of each
individual line scan (each $2 \ {\rm mm}$ long).

\begin{figure}
        \includegraphics[width=0.4\textwidth]{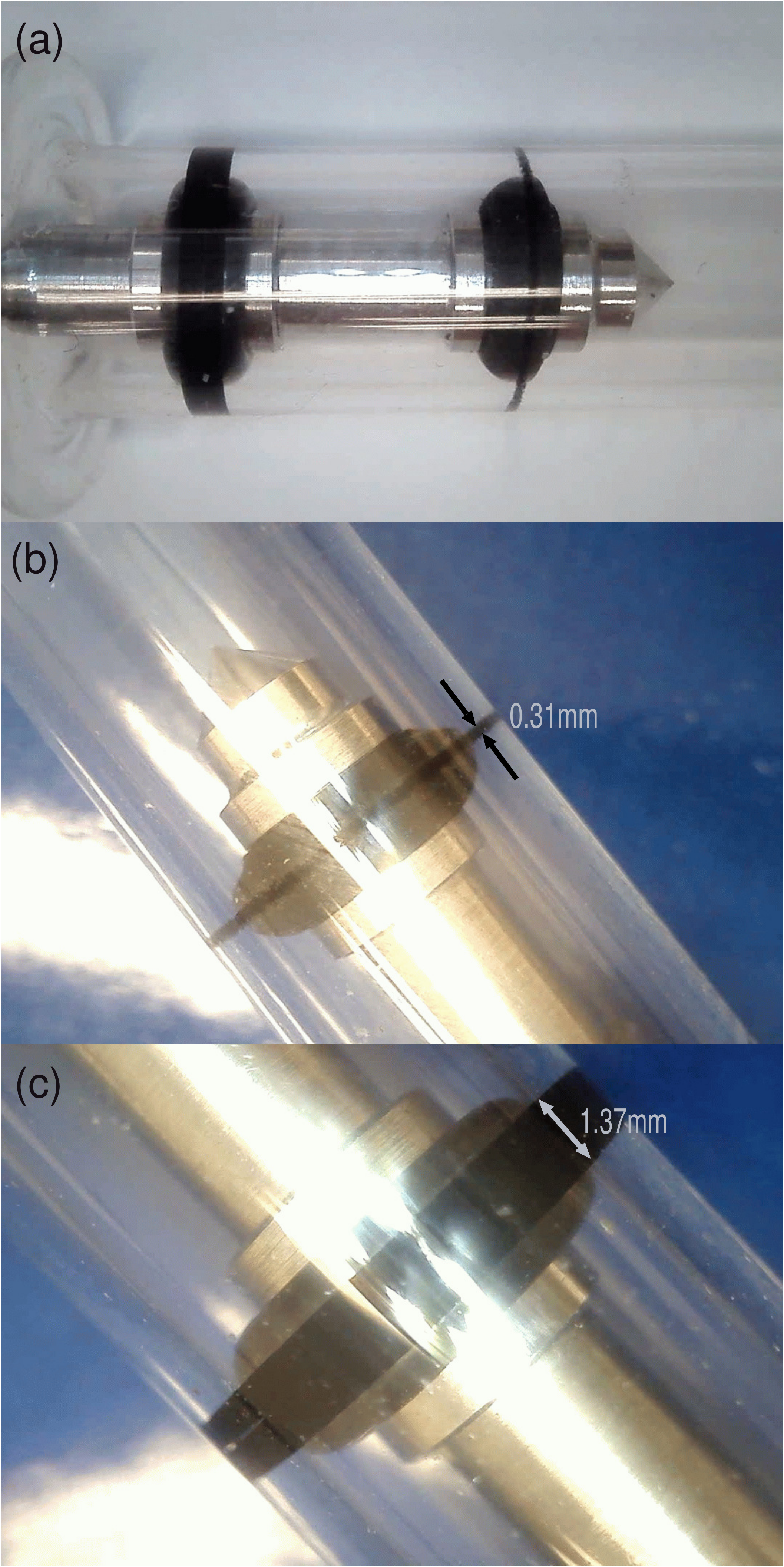}
        \caption{\label{BothInGlass.eps}
Optical pictures of two NBR70 rubber O-rings on a steel stopper inserted in a 
glass barrel. The front O-ring is sandblasted which reduces the outer radius of the
O-ring and results in a reduced width of the rubber-glass nominal contact region.
(a) shows both O-rings in the glass barrel and (b) and (c) the contact with the
sandblasted and not sandblasted O-ring, respectively.
}
\end{figure}

\begin{figure}
        \includegraphics[width=0.4\textwidth]{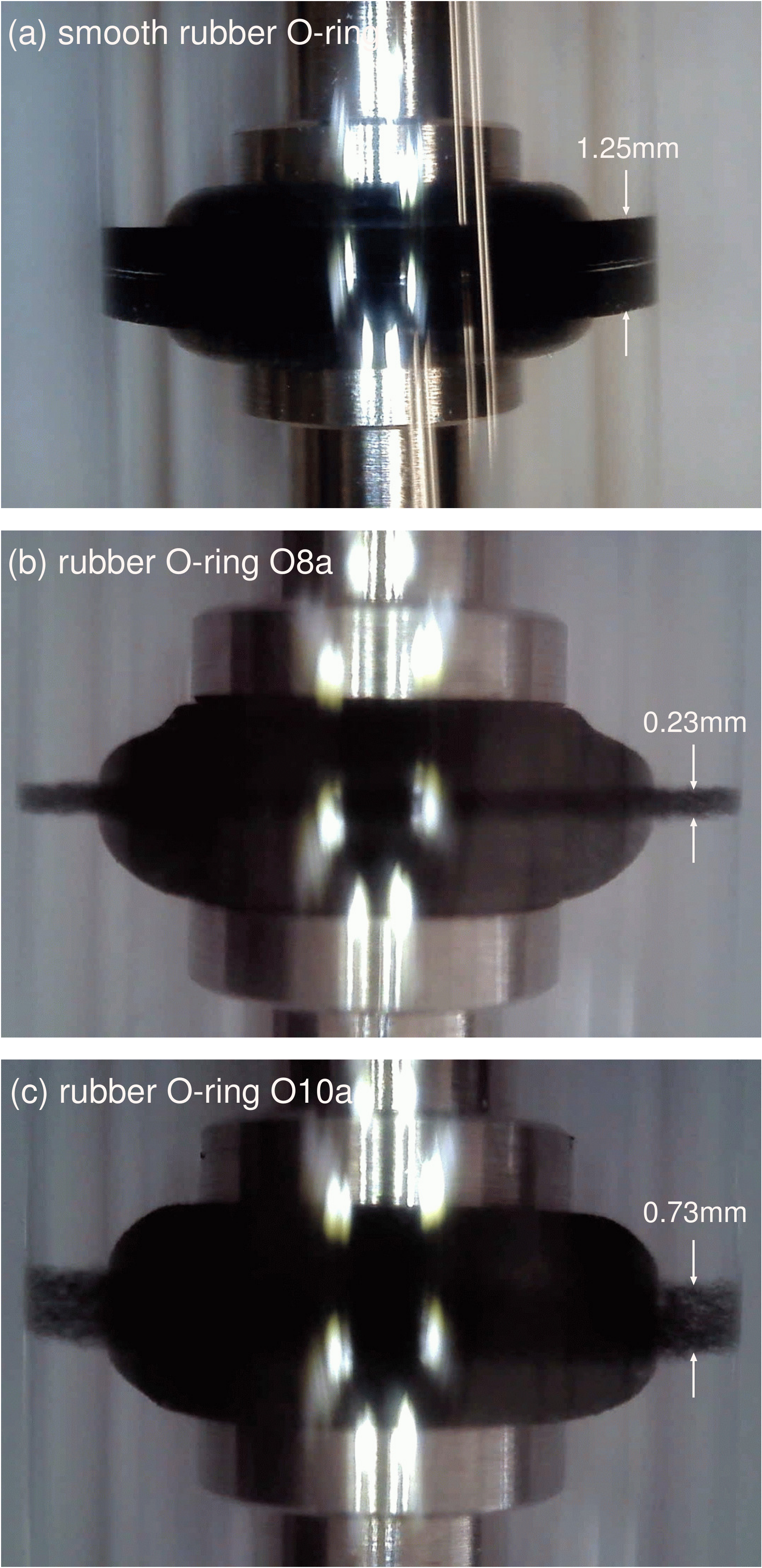}
        \caption{\label{RIBCONTACT1.eps}
	Optical picture of (a) a not sandblasted (smooth) O-ring and of
	sandblasted O-ring O8a (b) and O10a (c) in the glass barrel. 
	The width of the contact region are $w\approx 1.25$, $0.23$ and $0.73 \ {\rm mm}$, respectively.
	}
\end{figure}

\begin{table}[hbt]
	\caption{The measured contact width $w$ and the 
	calculated penetration $\delta$, normal force per unit length $F/L$,
	maximum contact pressure $p_0$, and the strain in the contact region
	for the studied O-rings. 
}
   \label{tab:oring1}
   \begin{center}
      \begin{tabular}{@{}|l||c|c|c|c|c|@{}}
         \hline
	      O-ring  & $w$ (mm) &  $ \delta $  (${\rm m m}$) & $F/L$ (N/m)  & $p_0$ (MPa) & strain (\%)  \\
         \hline
         \hline
            smooth & 1.25 & 0.391 & 2863 & 2.92 & 63 \\
         \hline
	    O1  & 0.54 & 0.073 & 534 & 1.26 & 27  \\
         \hline
	    O2  & 0.33 & 0.027 & 200 & 0.77 & 17  \\
         \hline
            O3  & 0.44 & 0.048 & 355 & 1.03 & 22 \\
         \hline
            O4  & 0.26 & 0.017 & 124 & 0.61 & 13 \\
         \hline
            O5  & 0.75 & 0.141 & 1031 & 1.75 & 38 \\
         \hline
	    O3a  & 0.50 & 0.063 & 458 & 1.17 & 25  \\
         \hline
	    O6a  & 0.69 & 0.119 & 873 & 1.61 & 35  \\
         \hline
            O8a  & 0.23 & 0.013 & 97 & 0.54 & 12 \\
         \hline
            O9a  & 0.40 & 0.040 & 293 & 0.93 & 20 \\
         \hline
            O10a  & 0.73 & 0.133 & 977 & 1.69 & 36 \\
         \hline
      \end{tabular}
   \end{center}
\end{table}

\vskip 0.2cm
{\bf 4.3 Optical pictures of the nominal rubber-barrel contact region}

The sandblasting results in a decrease in the effective O-ring radius and hence a change in the width of the Hertz-like
contact between the O-ring on the steel plunger and the glass barrel. This is illustrated in Fig. \ref{BothInGlass.eps}(c)
for a not sandblasted rubber O-ring, and in Fig. \ref{BothInGlass.eps}(b) for a strongly sandblasted O-ring. 
The width of the nominal contact area for the sandblasted O-ring is much smaller than for the not sandblasted 
O-ring ($0.31 \ {\rm mm}$ and $1.37 \ {\rm mm}$, respectively). In the experiments presented below the outer (not sandlasted)
O-ring was cut with a scalpel in such a way that a thin open channel occurred in the rubber-glass nominal contact region.
Hence, the whole pressure drop in the leakage experiments occurs over the inner sandblasted O-ring. 

We have measured the width of the rubber-glass contact region for all the O-rings used in the leakage studies. The measurements
was performed both in J\"ulich and at Sanofi using two different optical microscope. Fig. \ref{RIBCONTACT1.eps}
shows the results obtained in J\"ulich for rubber O-rings O8a and O10a and for a not sandblasted O-ring. In table 
\ref{tab:oring1} we show the width $w$ of the contact region for all the studied O-rings
as measured in J\"ulich. Using the Hertz theory for cylinder contact,
from the width of the contact region we can calculate 
the penetration $\delta = w^2/(4R)$, the loading force per unit length $F/L= (\pi/4) E^*\delta$,
the maximum contact pressure $p_0 = (E^*F/\pi R L)^{1/2}$ and the strain $2 \delta /w$.
In the calculations we have used $R=1 \ {\rm mm}$ and $E_1=E/(1-\nu^2)$
with $E=7 \ {\rm MPa}$ and $\nu = 0.5$. In calculation of the leakage rates presented below 
we have used (5) and assumed that the pressure distribution
$p(x)$ is of the Hertz type (4) with the width $L_x=w$ and the maximum pressure $p_0$ given in Table \ref{tab:oring1}. 

\begin{figure}
        \includegraphics[width=0.4\textwidth]{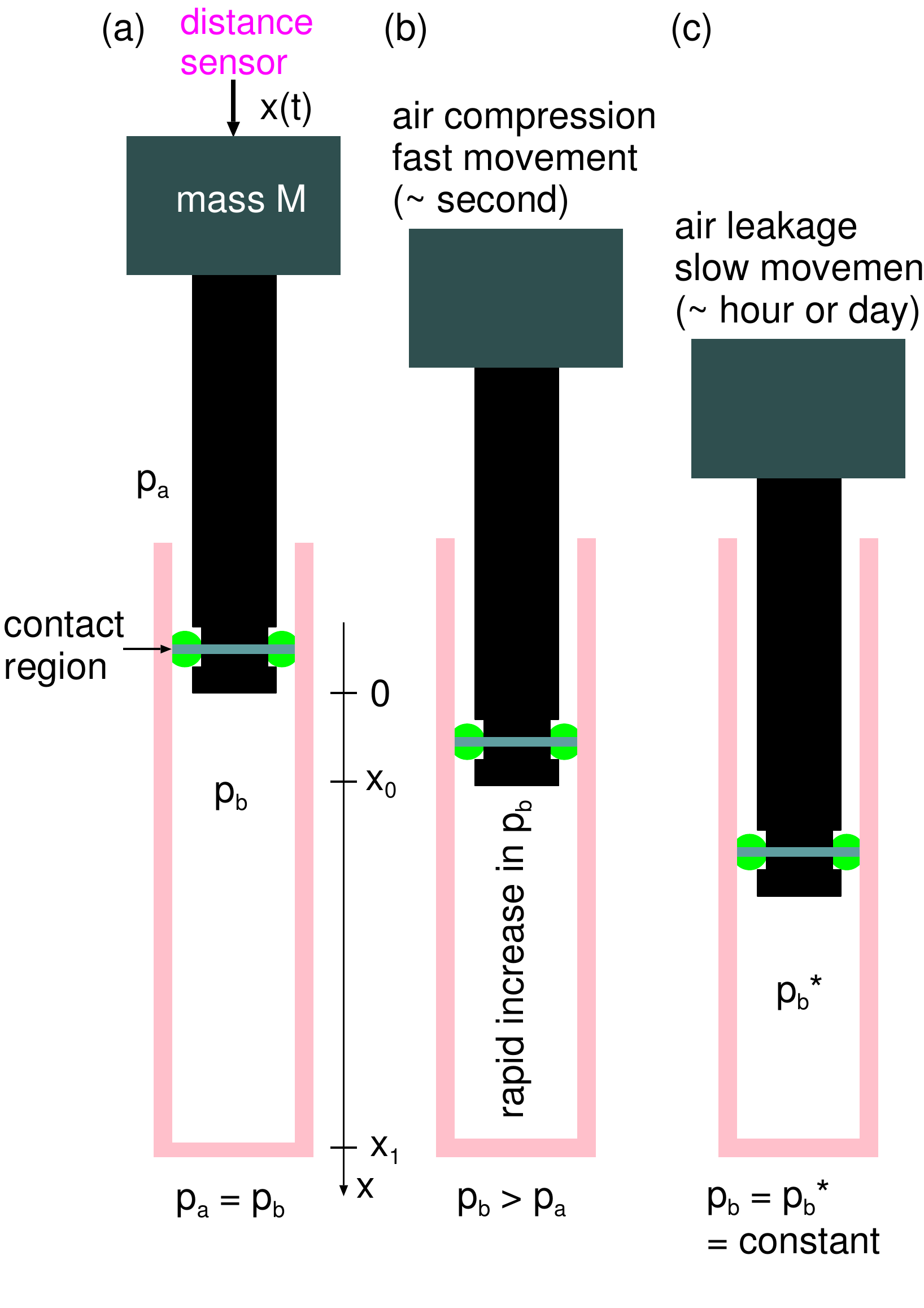}
        \caption{\label{JuelichLeakageExp.eps}
	The experimental set-up for measuring the air leakage rate (schematic).
	}
\end{figure}

\vskip 0.2cm
{\bf 4.4 Air leakage experiments and theory}

        The air leakage experiments was performed in J\"ulich using a home-made leakage set-up shown
	schematically in Fig. \ref{JuelichLeakageExp.eps}.
	In the actual experiments the steel plunger has two rubber O-rings (as in Fig. \ref{BothInGlass.eps}) 
	in order to give a more stable (stiff) set-up, but the second O-ring has 
	an axial cut (gas flow channel) and will not give any resistance to the gas or fluid flow but will contribute
	the to friction force acting on the plunger.

Let us apply the normal axial force $F_x=Mg$ (where $M$ is the loading mass) to the plunger
at time $t=0$ and assume all the air is removed
at $t=t_1$, where the stopper has moved to the bottom $x=x_1$ of the barrel (i.e., reached the full air squeeze-out position)
(see Fig. \ref{JuelichLeakageExp.eps}).
The leakage experiment involves two phases. In the initial phase, from time $t=0$ to $t=t_0$, 
with typically $t_0 \approx 10 \ {\rm sec}$, 
the stopper moves rapidly while compressing the air in the barrel.
During this act the air leakage can be neglected so the number of gas molecules
in the barrel equal
$$N_0 ={p_{\rm a} V_0 \over k_{\rm B}T} , \eqno(31)$$
where $V_0$ is the initial volume of trapped air at atmospheric pressure $p_{\rm a}$.
During this rapid motion the air pressure in the barrel increases
from $p_{\rm a}$ (the external pressure) to a higher pressure $p_{\rm b}^*$ while the gas volume
change from $V_0$ to $V_{\rm b}^*$ given by 
$$V_{\rm b}^*=V(t_0)=V_0-x_0 A_0,$$
where $x_0=x(t_0)$.
The pressure $p_{\rm b}^*$ is assumed to be given by the ideal gas law
$$p_{\rm b}^* V_{\rm b}^* = N_0 k_{\rm B}T . \eqno(32)$$

Assume that the properties of the glass barrel does not change with the lateral position
$x$. In this case, after the initial fast compression of the gas, the pressure will remain 
at the value $p_{\rm b}^*$ while the stopper moves into the barrel with a constant speed $\dot x$
squeezing out the air from the barrel. The air leakage can be calculated by dividing the number of gas molecules
removed, which equals $N_0$, by the time it takes, which equals $\Delta t = t_1-t_0$, giving
$$\dot N \approx {N_0 \over \Delta t}.\eqno(33)$$ 
We can also calculate the friction force during the slow movement of the stopper using
$$F_{\rm f}=F_x-\left (p_{\rm b}^*-p_{\rm a} \right) A_0.\eqno(34)$$

We have found that usually the properties of the glass barrel varies slightly with the lateral position
$x$. This may result from and $x$-dependency of the internal diameter of the glass barrel, or from variations
in the rubber-glass frictional properties with $x$. 
This will result in a time variation in the speed $\dot x$ of the stopper (or O-ring). In this case the 
gas pressure inside the barrel, and the friction force $F_{\rm f}$, will also depend on time (or $x$)
even when $t>t_0$. This problem can be studied analytically too, 
but in what follows we will just present the average leakage rate obtained using (33).

\begin{figure}
        \includegraphics[width=0.45\textwidth]{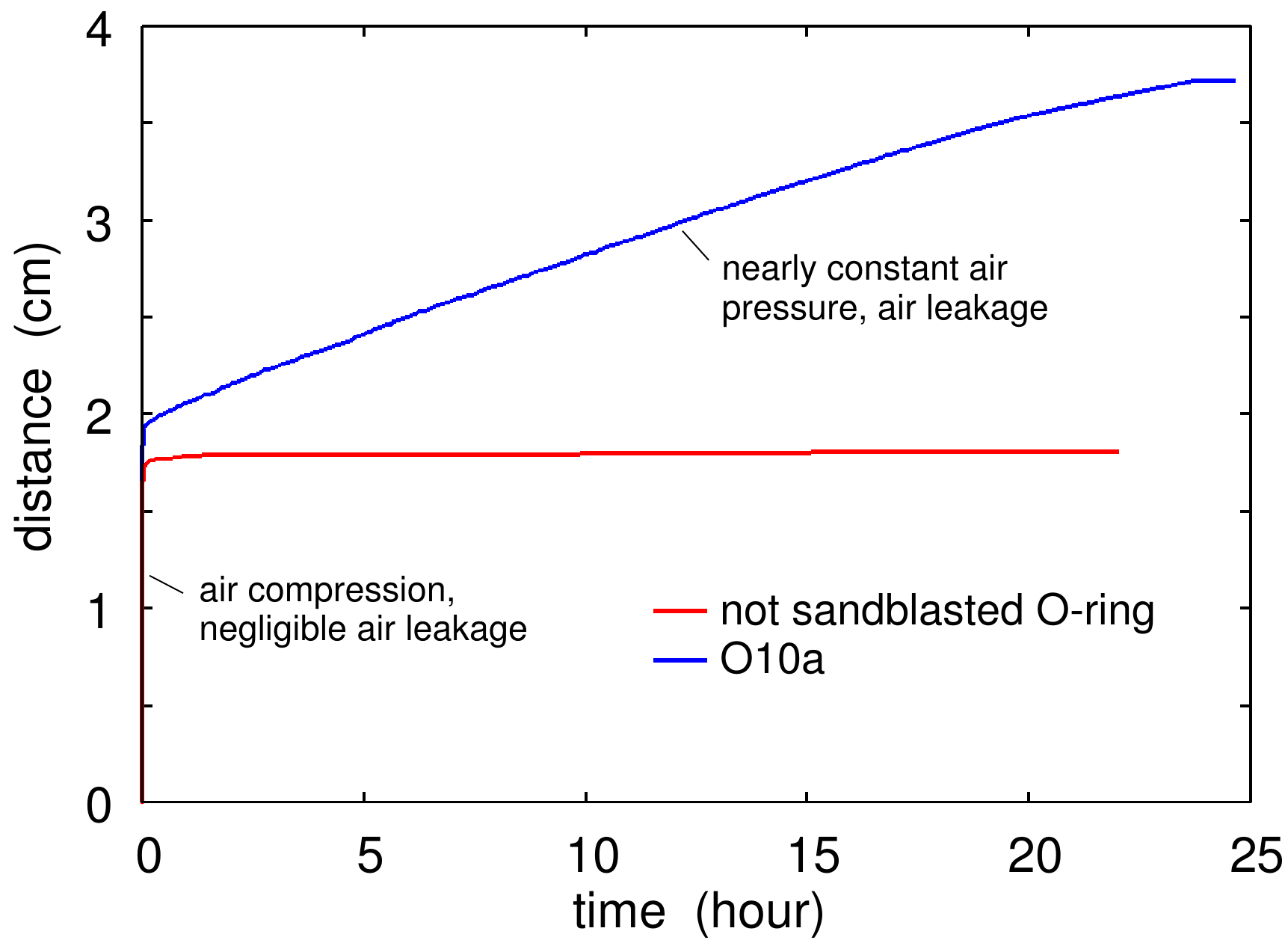}
        \caption{\label{1time.2distance.O3.O2.closed.needle.eps}
	The distance moved by the O-ring (or stopper) as a function of time for the O-ring
	O0 (red line) and O10a (blue line) with closed needle. The external (axial) force on the
	stopper is $F_x=4.4 \ {\rm N}$. 
}
\end{figure}

\begin{figure}
        \includegraphics[width=0.45\textwidth]{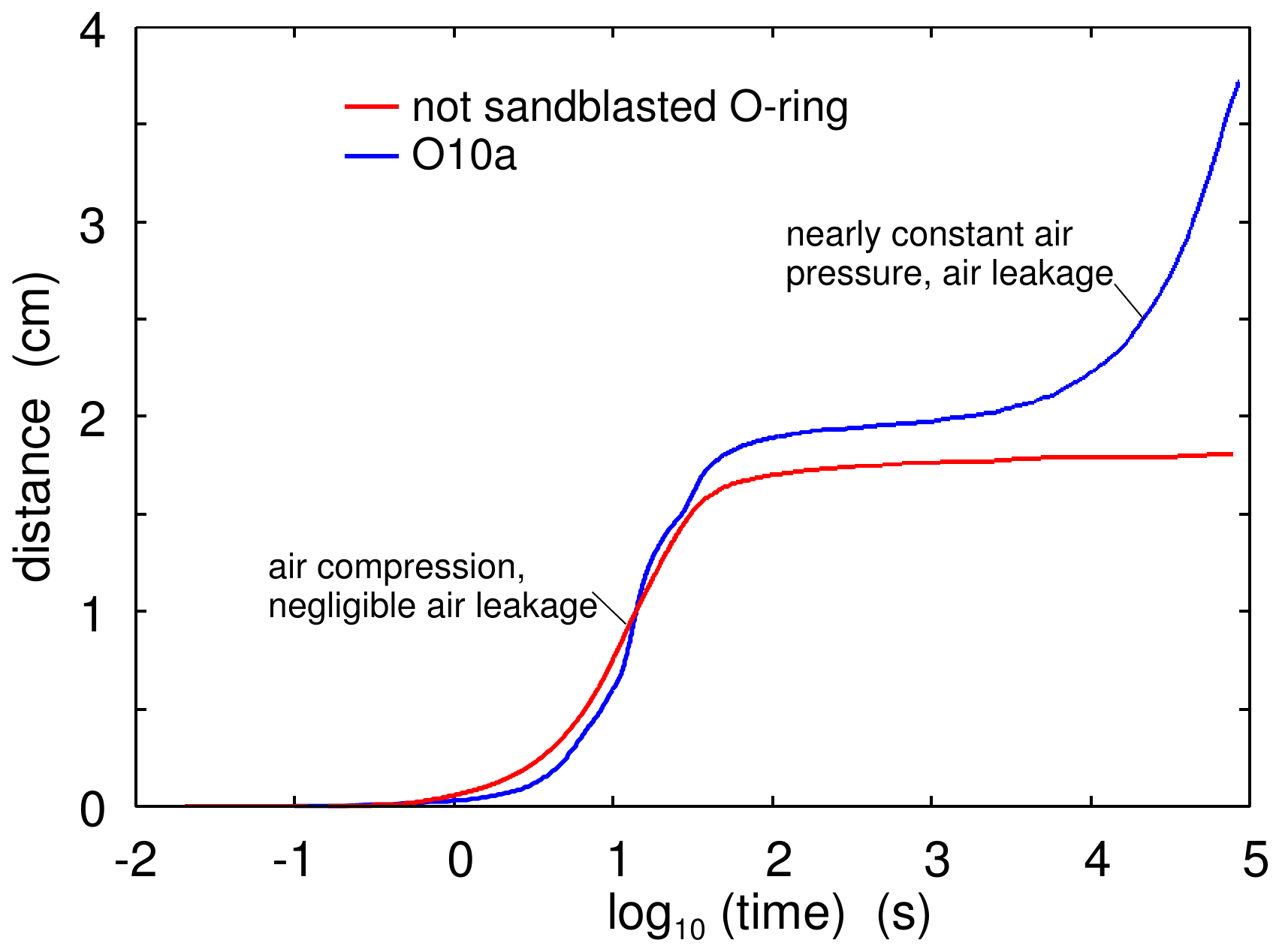}
	\caption{\label{1logTime.2x.O9a.and.NotUsed.eps}
	The distance moved for O-ring O10a (blue) and for a not sandblasted O-ring O0 (red),
as a function of the logarithm of time.}
\end{figure}

\begin{table}[hbt]
	\caption{
	The width of the rubber-glass contact region $w$ (in mm) in the air and He leakage experiments, and the measured
	He and air leakrate $\dot Q$ (in ${\rm cm^3/s}$ or equivalently ${\rm mbar} \times {\rm liter / s}$) for all studied O-rings. 
	For the He leakage the He pressure is 1 bar inside the syringe and  0 bar outside.
	For the air leakage for O-rings O1, O3-O5 the air pressure is $p_{\rm b}^*
	\approx 1.4 \ {\rm  bar}$ inside the syringe and  $1 \ {\rm  bar}$ outside.
	For the air leakage O-rings O2, O3a, O6a, O9a, O10a the air pressure is 2.2 bar inside the syringe and  1 bar outside.
        The glass barrels were cleaned to remove (most of) the silicone oil. For the not sandblasted O-ring O0 the air leakage
	experiment does not detect any leakage over a time period of 24 hours.
	The He-leakage experiment for the same system gave $5.1 \times 10^{-7}  \ {\rm cm^3/s}$
	which we do not attribute to true leakage at the rubber-barrel interface but to some other source of He atoms, e.g. desorption
	of He atoms from the camber wall. All He leakage rates below $10^{-6}  \ {\rm cm^3/s}$ are indicated by $0$ in the table 
	as they may be due to not to He originating from rubber-barrel interfacial He leakage.
}
   \label{tab:oring}
   \begin{center}
      \begin{tabular}{@{}|l||c|c|c|c|@{}}
         \hline
	      O-ring   & $w$ (air) &  $w$ (He) & $\dot Q$  (air) & $\dot Q$  (He) \\
	         &  (mm) & (mm)  & ${\rm cm}^3/s$ & ${\rm cm}^3/s$ \\
         \hline
         \hline
	      O0 & 1.25 & 1.11 & 0 & 0 \\
         \hline
	      O1 & 0.54 & .. & 0 & .. \\
         \hline
	      O2 & 0.33 & 0.35 & (0; 0) & 0  \\
         \hline
	      O3 & 0.44 & 0.49 & 0 & 0 \\
         \hline
	      O4 & 0.26 & 0.26 & $3.2\times 10^{-5}$ & $7.0 \times 10^{-5}$ \\
         \hline
	      O5 & 0.75 & 0.59 & 0 & (0; 0) \\
         \hline
	      O3a & 0.52 & 0.52 & (0; 0) & $1.0\times 10^{-5}$  \\
         \hline
	      O6a & 0.66 & 0.69 & 0  & $6.5\times 10^{-6}$ \\
         \hline
	      O8a  & 0.27 & 0.23 & $(3.3; 2.1;1.2)\times 10^{-3}$ & $5.0\times 10^{-3}$  \\
         \hline
	      O9a & 0.43 & 0.40 & 0  & 0 \\
         \hline
	      O10a & 0.61 & 0.72 & $1.3\times 10^{-5}$ & $1.2\times 10^{-5}$ \\
         \hline
      \end{tabular}
   \end{center}
\end{table}

\vskip 0.2cm
{\bf Experimental air leakage results}

We have studied the leakage of air for a large number of O-rings with different surface roughness
and different effective radius of the O-ring (which depends on the time of sandblasting) 
and hence different width of the rubber-barrel contact region (see Fig. \ref{BothInGlass.eps}, \ref{RIBCONTACT1.eps} for 
some optical pictures of the contact region). 
In the experiments we applied an external axial force $F_x$ to the plunger which for the O-rings O1, O3-O5 was
$F_x = 2.2 \ {\rm N}$ giving the pressure in the compressed
air $p_b^* \approx 1.4\ {\rm bar}$ 
as determined by the volume change during the initial fast air compression act (see Fig. \ref{BothInGlass.eps}(b)).
For the O-rings O0, O2, O3a, O6a, O8a, O9a, O10a we used a bigger axial force $F_x= 4.4\ {\rm N}$ giving the compressed air pressure
$p_b^* \approx 2.2\ {\rm bar}$. 
We note that in all cases the friction force $F_{\rm f}$ during the slow motion (typical velocity $\approx 1-10 \ {\rm \mu m/s}$)
during the air squeeze-out, is about $1 \ {\rm N}$, and is mainly determined by the second (not sandblasted) O-ring 
(with an axial cut), which was the same in all the experiments.

Depending on the surface roughness and the contact pressure we observe
that most of the O-rings do not leak (on the time scale of $\sim 20 \ {\rm hour}$) while O-rings O4, O8a and O10a leak (complete
removal of the air in $\sim 1-20 \ {\rm hour}$). To illustrate this, in Fig. \ref{1time.2distance.O3.O2.closed.needle.eps}
we show the distance $x$ moved by the O-ring (or stopper) as a function of time for the not sandblasted O-ring
O0 (red line) and O10a (blue line) with closed needle. For the O-ring O0, no air leakage occurred and the movement of the
stopper (about $2 \ {\rm cm}$) resulted in a compression of the air in the syringe so that 
the air pressure force $p_b^* A_0$, and the friction force $F_{\rm f}$, just balance the applied force:
$F_x=F_{\rm f}+p_b^*A_0$ ($A_0$ is the inner barrel cross section). 

Fig. \ref{1logTime.2x.O9a.and.NotUsed.eps} shows similar results on a logarithmic time scale 
for a not sandblasted O-ring O0 (red line) and for O-ring O10a (blue line). For O-ring O10a the air pressure $p_{\rm b}^*$ in the barrel
is $2.2 \ {\rm bar}$ and the air leakrate $\dot V = 1.3\times 10^{-5} \ {\rm cm^3/s}$. The total
measurement time in this experiment was $\approx 24 \ {\rm hours}$. 

In Table \ref{tab:oring} we summarize the results of all the measured air leakage rates, $\dot Q$ in ${\rm cm^3/s}$, 
or equivalently ${\rm mbar} \times {\rm liter / s}$.
We also give the width of the rubber-glass contact region $w$ (in mm), and the measured
He leakrates for all the studied O-rings. 
For the air leakage for O-rings O1, O3-O5 the air pressure is $\approx 1.4 \ {\rm  bar}$ inside the syringe and  $1 \ {\rm  bar}$ outside,
while for the O-rings O0, O2, O3a, O6a, O9a, O10a the air pressure is 2.2 bar inside the syringe and  1 bar outside.
The glass barrels were cleaned to remove 
the silicone oil by immersing them in a bath of heptane for 12 hours, after which they where dried for a couple of days.
For the O-rings
where the leakage rates are indicated as $0$, after the fast compression which occurs in typically $\sim 10 \ {\rm s}$,
we could not detect any movement of the stopper during a time period of
at least $\sim 10 \ {\rm hours}$.

\begin{figure}
        \includegraphics[width=0.45\textwidth]{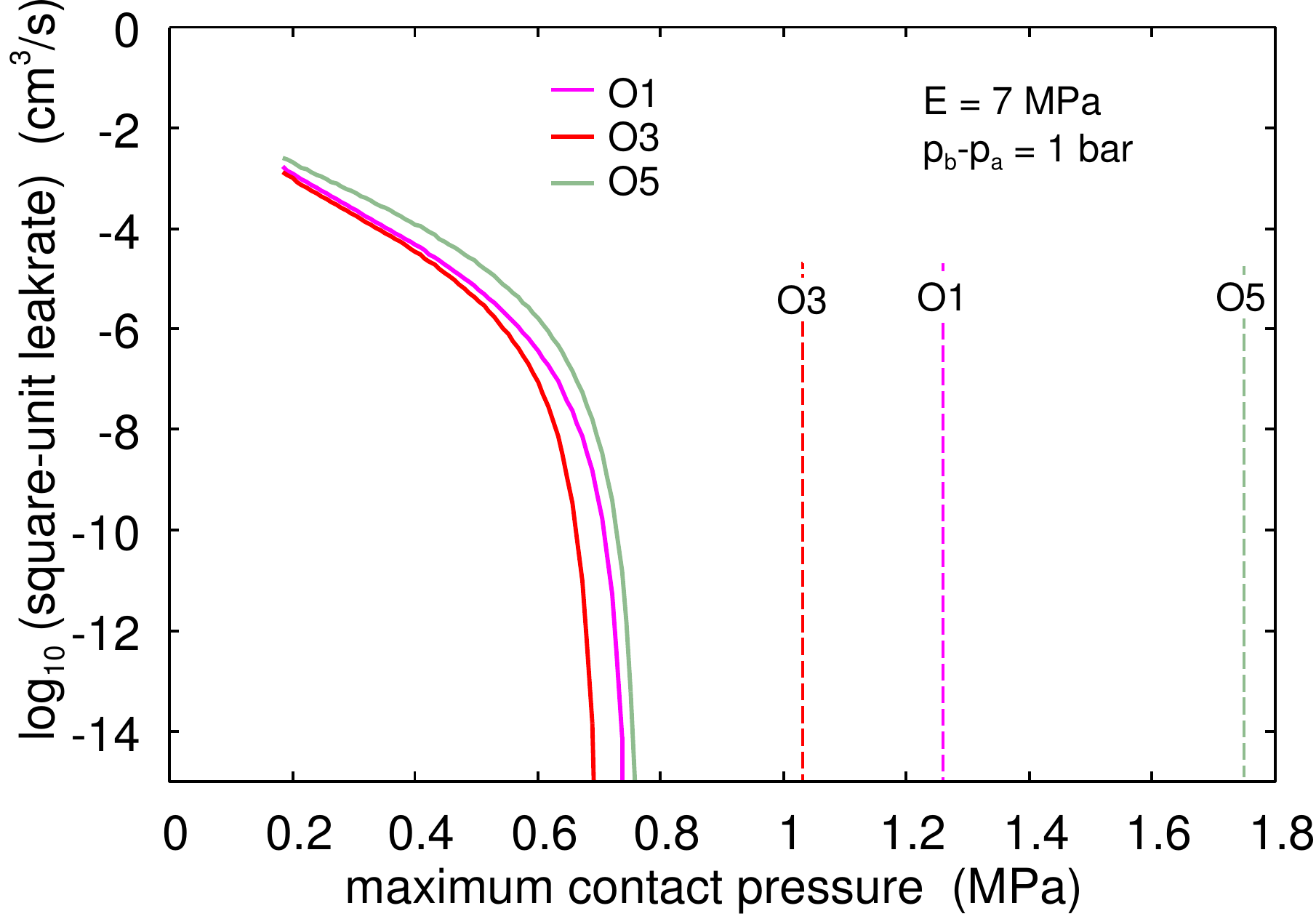}
        \caption{\label{1pcontact.2logLeakage.O3.E=6MPa.eps}
	The calculated square-unit leakrate for O-ring O1 (pink), O3 (red) and O5 (gray)
	as a function of the maximum pressure in the Hertz contact region.
	The vertical dashed lines 
	indicate the actual maximum pressures 
	($p_0\approx 1.26$, $1.03$ and $1.75 \ {\rm MPa}$, respectively) 
	in the rubber-glass contact region.
        The theory predict no gas leakage for these O-rings which agree with observations.
	}
\end{figure}

\begin{figure}
        \includegraphics[width=0.45\textwidth]{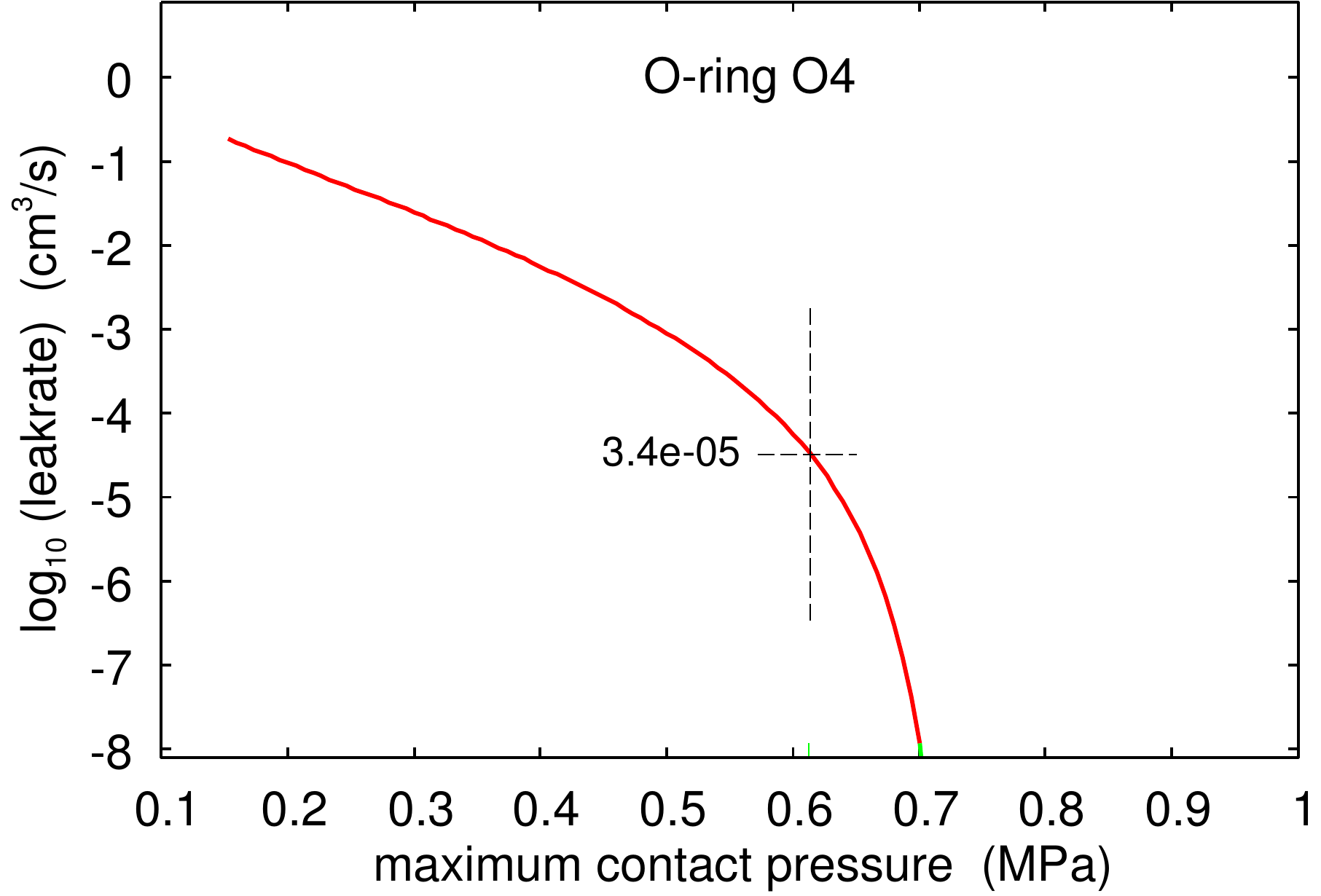}
        \caption{\label{1p0.2leakage.O2.O4.combined.eps}
	The calculated leakrate $\dot Q$ for O-ring O4  
	as a function of the maximum pressure $p_0$ in the Hertz contact region.
	}
\end{figure}

\begin{figure}
        \includegraphics[width=0.45\textwidth]{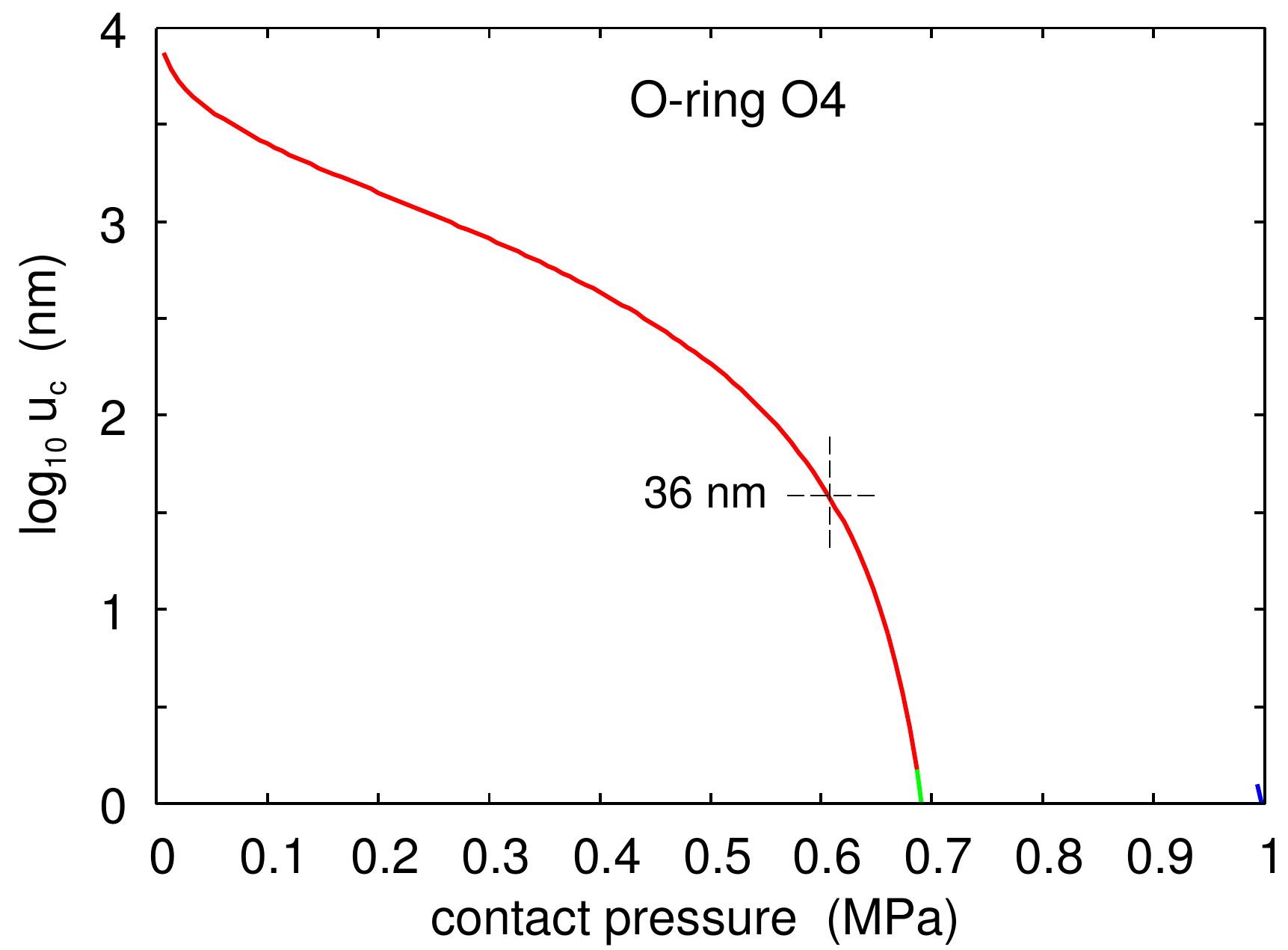}
        \caption{\label{1pressure.2uc.E=7MPa.O2.O4.eps}
	The surface separation $u_{\rm c}$ at the critical constrictions for O-ring O4
	as a function of the maximum pressure $p_0$ in the Hertz contact region.
	The vertical dashed lines 
	indicate the actual maximum pressures $p_0 \approx 0.61 \ {\rm MPa}$
	in the rubber-glass cintact region,
	as calculated using the Hertz theory (see Table. \ref{tab:oring1}).
	}
\end{figure}

\begin{figure}
        \includegraphics[width=0.45\textwidth]{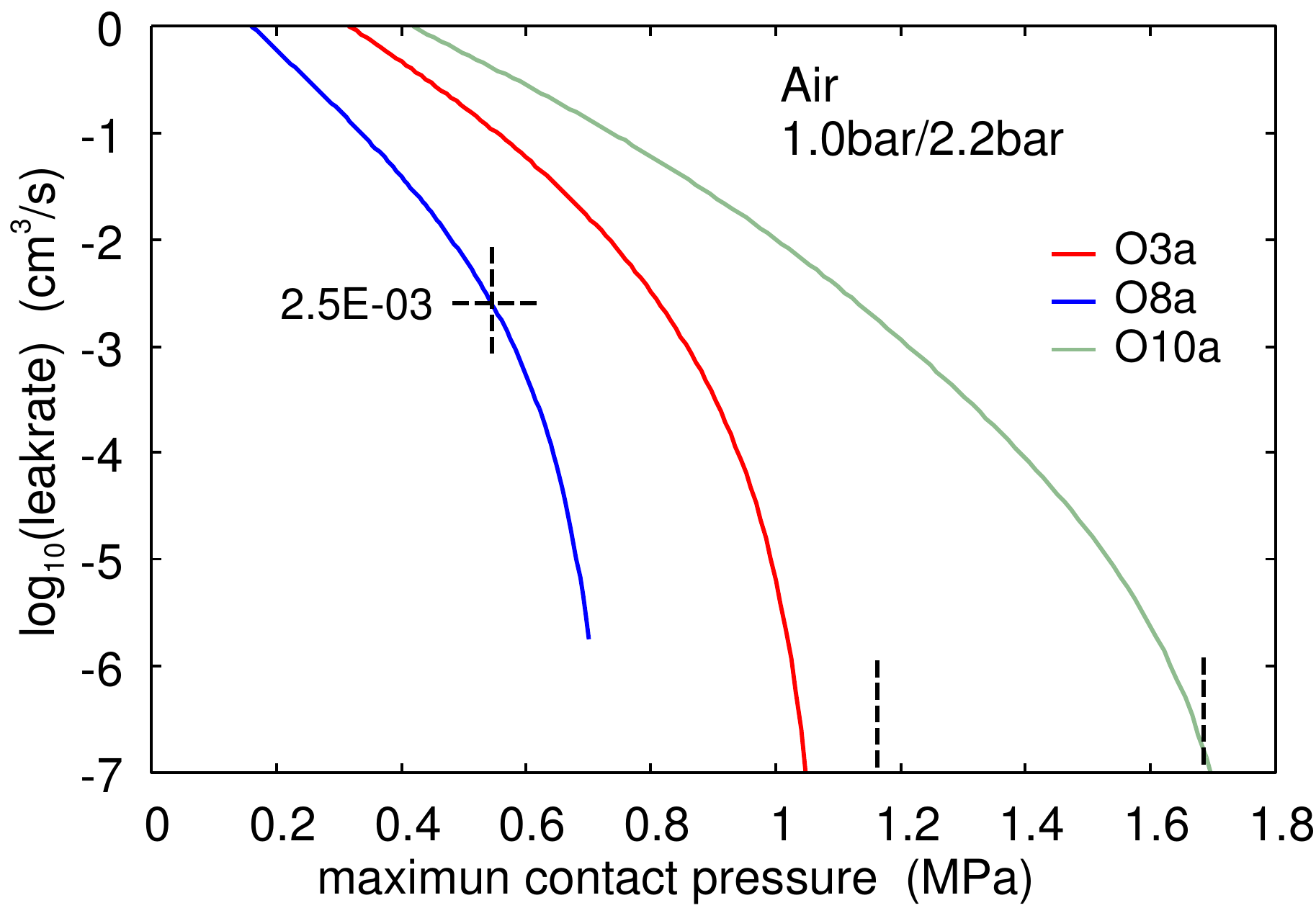}
        \caption{\label{1pressure.2logQdot.O3a.O8a.O10a.2.2barInsideAir.corrected.eps}
	The logarithm of the calculated air leakage rate as a function of the maximum pressure in the 
	Hertz contact pressure distribution for O-rings O3a, O8a and O10a. 
        In the experiment the air pressure in the syringes is $2.2 \ {\rm bar}$ and outside $1 \ {\rm bar}$.
	}
\end{figure}

\begin{figure}
        \includegraphics[width=0.45\textwidth]{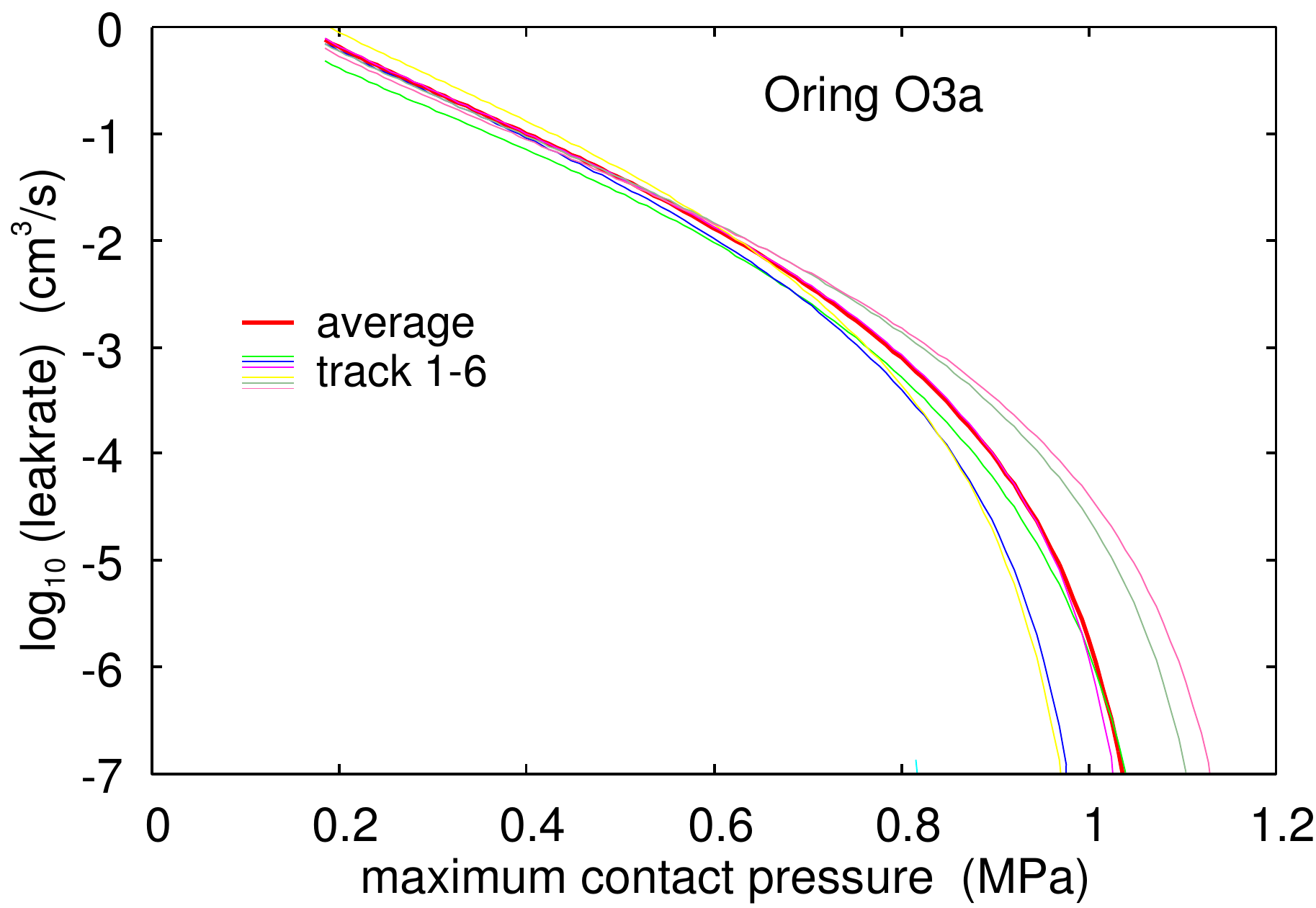}
        \caption{\label{1pressure.2logLeakage.O3a.many.eps}
	The logarithm of the calculated air leakage rate as a function of the maximum pressure in the 
	Hertz contact pressure distribution for O-ring O3a. Results are shown using the power spectra of
	7 line tracks each 2 mm long (thin lines) and using the average power spectrum
	(thick line) shown in Fig. \ref{1logq.2logC2D.O3a.all1.eps}. 
	}
\end{figure}

\begin{figure}
        \includegraphics[width=0.45\textwidth]{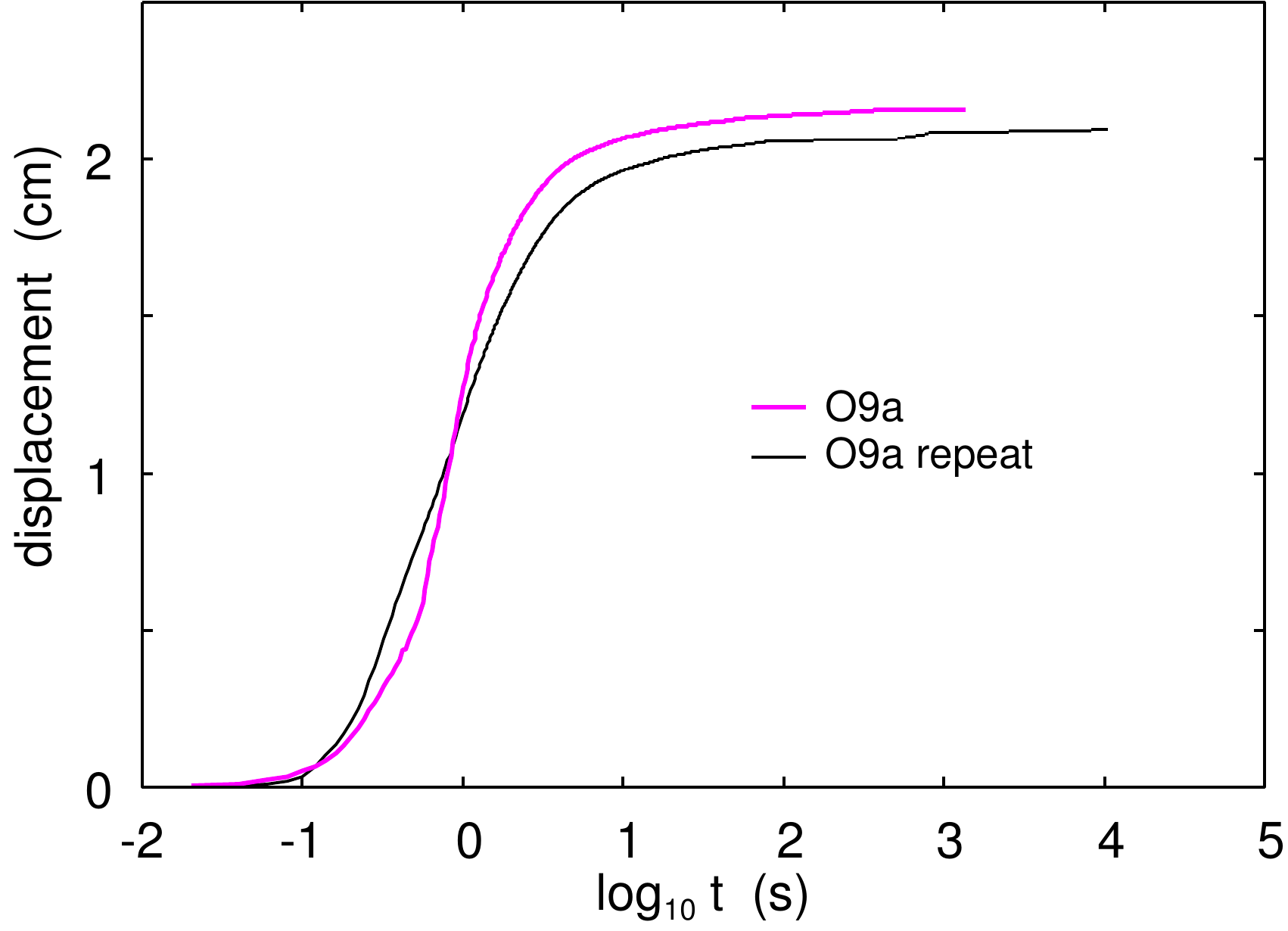}
	\caption{\label{1logTime.2x.9a.repeat.long.3.eps}
	The displacement of  the rubber stopper in measurements for O-ring O9a 
	in two different glass barrels from the same batch.
	}
\end{figure}

\vskip 0.2cm
{\bf Analysis of the air leakage experiments}

Using the measured surface roughness power spectra we have calculated the air leakage rates
for all the O-rings. In the calculations we have assumed the Hertz contact pressure (4) with the
maximum pressure $p_0$ and the width $w$ given in Table \ref{tab:oring1}. In the calculation of the 
pressure $p_0$ we have used the rubber modulus $E=7 \ {\rm MPa}$ (and Poisson ratio $\nu = 0.5$)
but for a linear viscoelastic material 
the calculated leakrate is in fact independent of the $E$-modulus because 
the interfacial separation $u(x,y)$ does
not depend on the $E$-modulus when the compression $\delta$ is given, as is the case for an O-ring confined inside
the barrel. Thus, the width $w$ is determined by the the radius $R$ and the compression $\delta$ (as $w=(4R\delta)^{1/2}$) 
and does not depend on the $E$-modulus. The maximum pressure $p_0$ does increase linearly 
with increasing $E$-modulus but the separation at the interface depends on $p/E$ 
and is hence independent of the $E$-modulus. For real rubber materials, which exhibit strain softening for small strain (and strain-stiffening
for large strain), this results holds accurately only if the strain and hence the effective modulus is similar in the macroscopic
Hertz contact region as in the asperity contact region. The macroscopic strain is typically $10-60\%$ (see Table. \ref{tab:oring1}) while
the strain in the asperity contact regions may be slightly larger but for large strain the strain softening
does not vary strongly with the strain level.

Table \ref{tab:oring} shows that for O-rings O0-O3, O5, O3a, O6a and O9a we did not observe any air leakage. 
This is in agreement with our calculated leakrates. Thus in Fig. \ref{1pcontact.2logLeakage.O3.E=6MPa.eps}
we show the calculated square-unit leakrate for O-ring O1 (pink), O3 (red) and O5 (gray)
as a function of the maximum pressure in the Hertz contact region.
In the calculation we have used the measured surface roughness power spectra of the O-rings. 
The vertical dashed lines indicate the actual maximum pressures 
$p_0\approx 1.26$, $1.03$ and $1.75 \ {\rm MPa}$, respectively (see Table \ref{tab:oring1}), obtained from the width $w$
of the contact region.
In the calculation we have assumed that the air pressure in the syringe is $2 \ {\rm bar}$
and $1 \ {\rm bar}$ outside the syringe, but the leakage for any other pressure difference $\Delta p$ can be obtained
by direct scaling since the ballistic contribution to the leakrate is proportional to $\Delta p$. 
Clearly, at the maximum pressure in the Hertz contact region the contact area percolate in all
cases, and the leakage rate vanish. 

For the O-ring O4 we do observe leakage with $\dot Q =3.2 \times 10^{-5} \ {\rm cm^3/s}$.
In Fig. \ref{1p0.2leakage.O2.O4.combined.eps}
we show the calculated leakrate $\dot Q$ for O-ring O4 (green line) 
as a function of the maximum pressure $p_0$ in the Hertz contact region.
In the calculation we have used the measured surface roughness power spectrum of the O-ring O4
(the 1D power spectra are shown in Fig. \ref{1logq.2logC1D.O1O2O3.good.eps}, and was converted into the 2D power spectra assuming
isotropic roughness). The vertical dashed line 
indicate the actual maximum pressures $p_0 \approx 0.61 \ {\rm MPa}$
in the contact region for O-ring O4 in the glass barrel, 
as calculated using the Hertz theory (see Table \ref{tab:oring1}).
In the calculation we have used the (measured) air pressure in the 
syringes, namely $1.52 \ {\rm bar}$ for O-ring O4, 
and $1 \ {\rm bar}$ outside the syringe.
The leakrate assuming this pressure is
$\approx 3.4\times 10^{-5}  \ {\rm mbar \times liter/s}$ for O-ring O4,
in good agreement with the measured leakrate.

In Fig. \ref{1pressure.2uc.E=7MPa.O2.O4.eps}
we show the surface separation $u_{\rm c}$ at the critical constrictions for O-ring O4 
as a function of the maximum pressure $p_0$ in the Hertz contact region.
The vertical dashed lines indicate again the actual maximum pressures $p_0 \approx 0.61 \ {\rm MPa}$
in the contact region for the O-ring O4 in the glass barrel. The surface separation $u_{\rm c}$ for 
the O4 O-ring is about two times smaller than the air molecule mean free path, and 
the air leakage is mainly ballistic. 

Fig. \ref{1pressure.2logQdot.O3a.O8a.O10a.2.2barInsideAir.corrected.eps}
	shows the logarithm of the calculated air leakage rate as a function of the maximum pressure in the 
	Hertz contact pressure distribution for O-rings O3a, O8a and O10a. 
	The vertical dashed lines indicate the contact pressures calculated from the width $w$ of the contact region
	($p_0=1.17$, $0.54$ and $1.69 \ {\rm MPa}$). For the O-ringe O8a and 03a the predicted leakage rates
	($2.5\times 10^{-3} \ {\rm cm^3/s}$ and $\approx 0$, respectively) 
	are close to the observed leakage rates $(1.2-3.3)\times 10^{-3} \ {\rm cm^3/s}$
	and $0$, respectively.
        For the O-ring O10a the predicted leakage rate is smaller than observed, but a relative small change in the
	contact pressure (from $1.69 \ {\rm MPa}$ to $\approx 1.5  \ {\rm MPa}$) would give agreement between the calculated
	and measured leakage rates.

The agreement found above between theory and experiment is very good, in particular taking into account that the width $w$
measured optically is not highly accurate due to blurred nature of the contact line (due to the surface roughness, 
the curved nature of the glass barrel). 
The surface roughness power spectrum used in the calculations has also some uncertainty.
To illustrate this, in Fig. \ref{1pressure.2logLeakage.O3a.many.eps}
we show the logarithm of the calculated air leakage rate as a function of the maximum 
Hertz contact pressure for O-ring O3a. Results are shown using the power spectra of
the 7 line tracks, each 2 mm long (thin lines) and using the average power spectrum
(thick line) shown in Fig. \ref{1logq.2logC2D.O3a.all1.eps}. 

We have found that using glass barrels from different batch can have a relative large influence on the leakage rate.
We attribute this to small changes in the internal diameter of the glass barrel. Because of the small compression
(or penetration) for some rubber O-rings, e.g., only $\delta = 17 \ {\rm \mu m}$ for the O-ring O4 (see Table
\ref{tab:oring1}), the variation in the internal diameter specified by the tolerance $\pm 100 \ {\rm \mu m}$
can result in non-contact in some case. 
However typical variation within a batch of produced syringes is much smaller
and using glass barrels from the same batch gives reproducible results.
This is illustrated in Fig. \ref{1logTime.2x.9a.repeat.long.3.eps}
which shows the displacement of  the rubber stopper in measurements for O-ring O9a in two different 
glass barrels from the same batch.

\begin{figure}
        \includegraphics[width=0.45\textwidth]{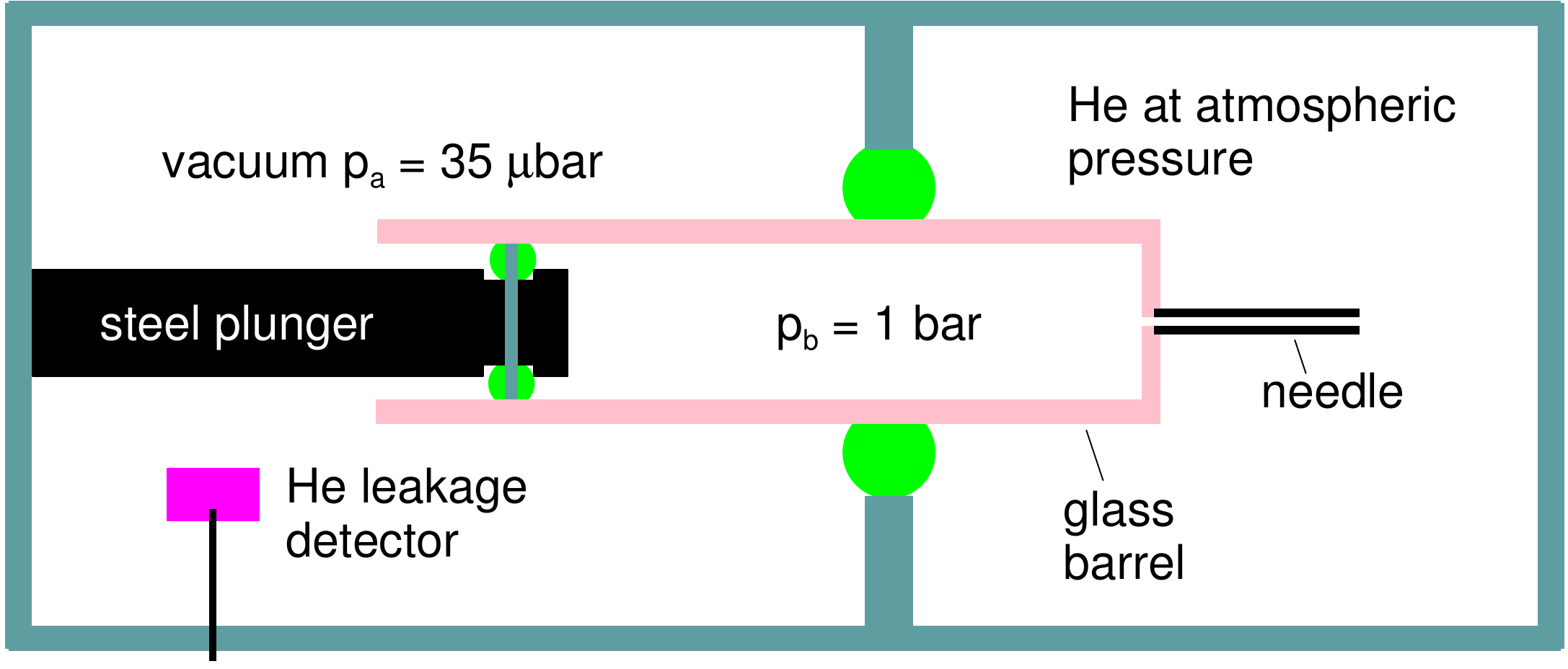}
        \caption{\label{SanofiHe.eps}
	Experimental set-up for measuring He gas leakage (schematic). 
	The He detector is a mass spectrometer.
	}
\end{figure}

\vskip 0.2cm
{\bf 4.5 Helium leakage experiments}

We have performed Helium (He) leakage experiments with the PFEIFFER Vacuum ASM340.
The schematic set-up is shown in Fig. \ref{SanofiHe.eps}.
The region outside of the syringe on the plunger side is located in a volume connected to a vacuum pump and 
a He leak detector.
The He detector (a mass spectrometer) monitors the He gas concentration.
After the syringe is placed in the chamber, in order to remove (or reduce) the He adsorbed on the
solid walls (which are in kinetic equilibrium with the He gas in the chamber) 
the vacuum pump keep a vacuum for at least 1 hour.
Once the chamber is decontaminated, He is introduced in the syringe, and the mass spectrometer detector
monitors the leakage of He through the rubber stopper. However, even for a not 
sandblasted rubber O-ring a finite He leak current is
observed at the rate 
$\dot Q ({\rm He}) \approx 5 \times 10^{-7} \ {\rm mbar \cdot liter/s}$ 
which must result from 
other He leakage sources, e.g., diffusion of He through the rubber or from He adsorbed
on the inner walls of the test equipment.
Hence, in the present set-up 
leakrates less than $\dot Q ({\rm He}) \approx  10^{-6} \ {\rm mbar \cdot liter/s}$
may occur even when there is no true interfacial He leakage.



In Table \ref{tab:oring} we summarize the results of all the measured leakage rates for He (and air). 
We also give the width of the rubber-glass contact region $w$. For the three O-rings O4, O8a and O10a
where we observe air leakage we observe similar He leakage rates. The He leakage rate for O-ring O4 and O8a
is about $\sim 2$ times higher than in air which may be due to the roughly $\sim 2.6$ times higher
average velocity $\bar v$ of He atoms at room temperature as compared to air molecules (note: the leakage rate is
proportional to $\bar v$ in the ballistic limit, see (30)). For O-rings O3, 05 and O9a the He leakage is below
$10^{-6} \ {\rm cm}^3$ which we do not attribute to interfacial glass leakage and with therefore are denoted
as zero leakage rate in Table \ref{tab:oring}, which agree with the air leakage observation. For the O-rings
O3a and O6a we observe no air leakage but the He leakage rates are of order $10^{-5}  \ {\rm cm}^3$. 

\begin{figure}
        \includegraphics[width=0.45\textwidth]{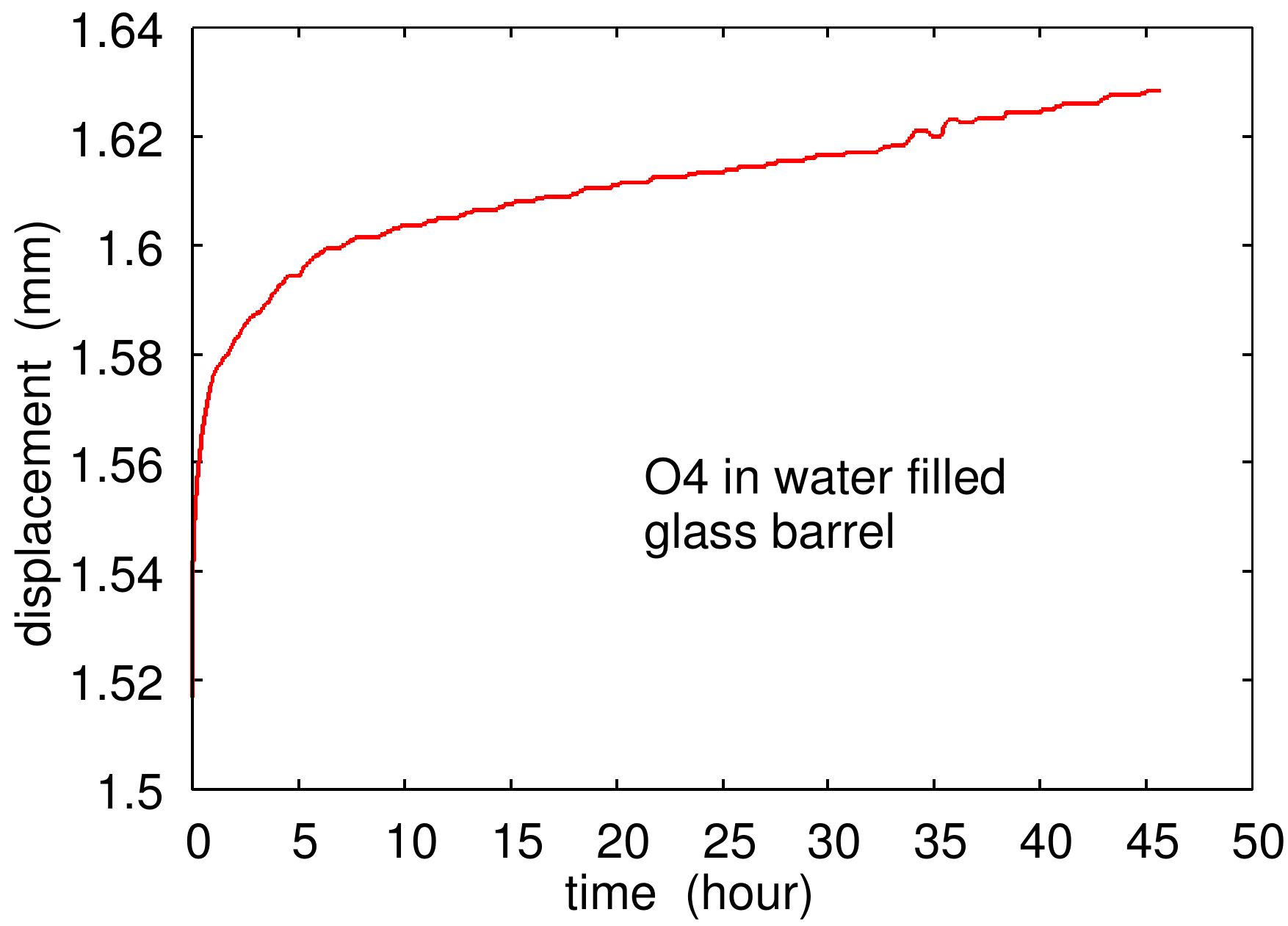}
        \caption{\label{1time.2x.mm.O2.second.barrel.in.water.eps}
The displacement of the stopper as a function of time for a water filled glass barrel for O-ring O4.
	}
\end{figure}

\begin{figure}
	\includegraphics[width=0.4\textwidth]{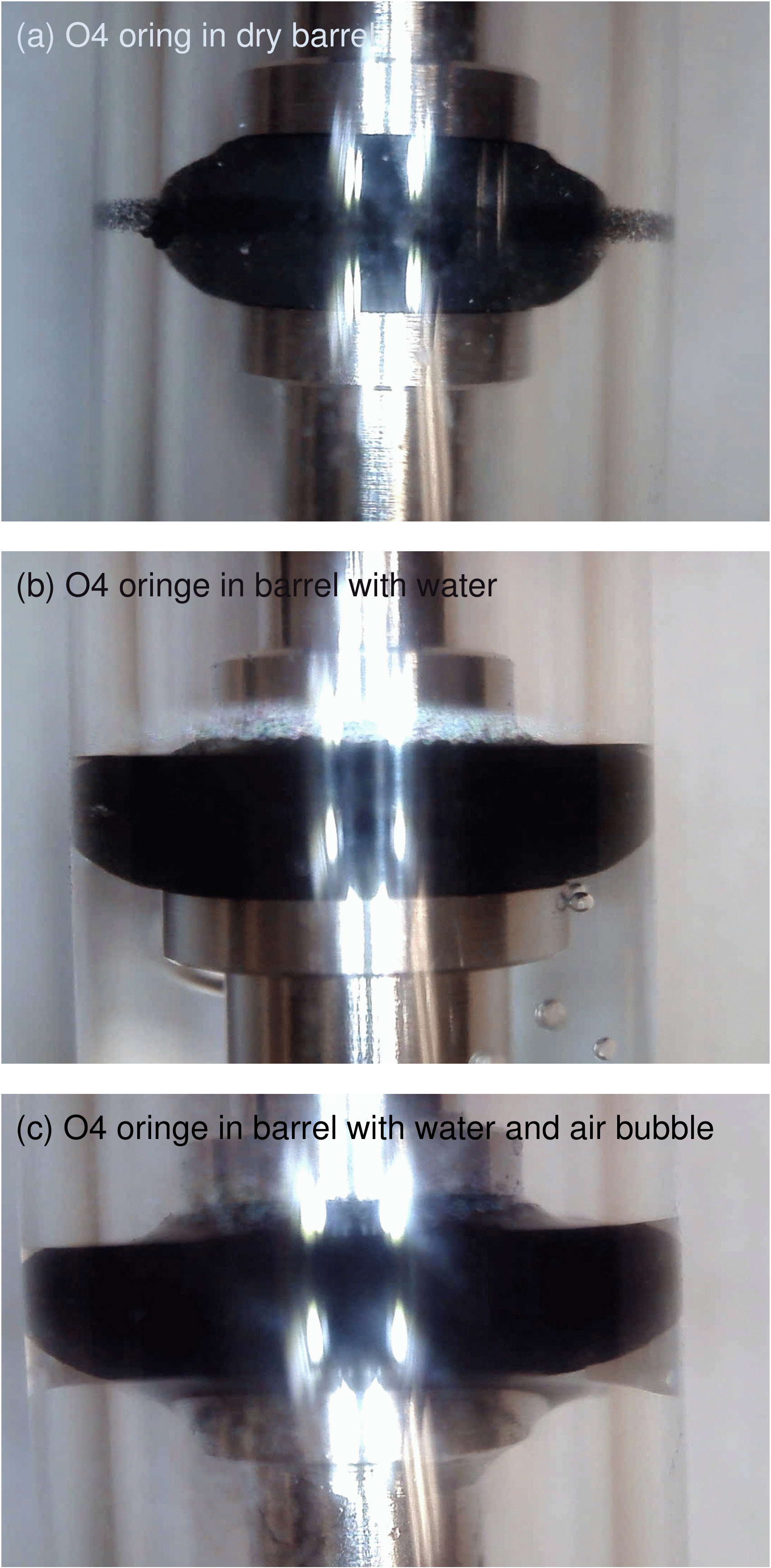}
        \caption{\label{WEToringe.eps}
	O-ring O4 in (a) dry glass barrel, (b) water filled barrel and (c) water filled barrel with air bubble
	close to the O-ring. In case (c) a water capillary bridge exist at the O-ring.
	}
\end{figure}

\begin{figure}
	\includegraphics[width=0.3\textwidth]{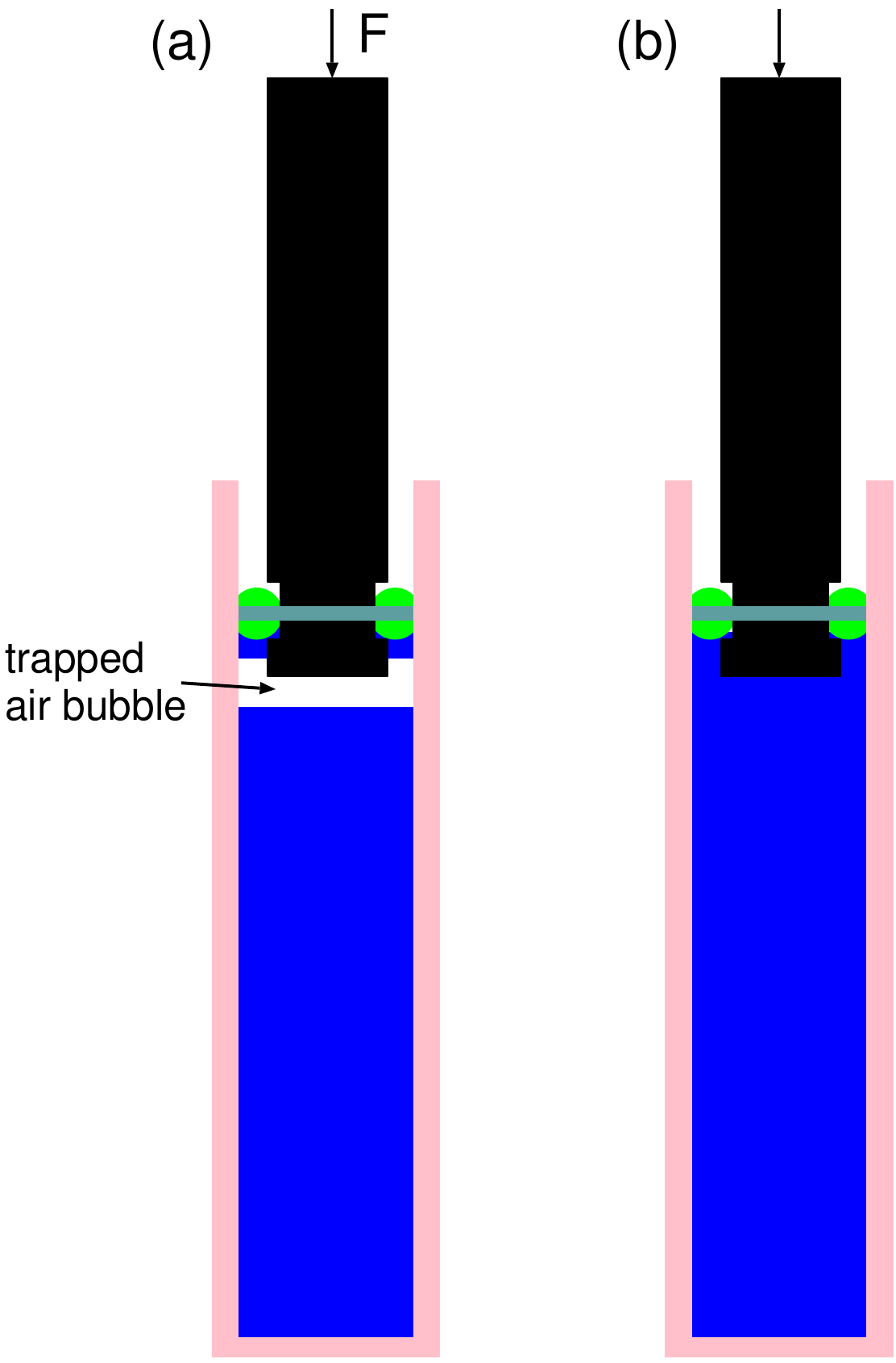}
        \caption{\label{CapilaryPic.eps}
	Glass barrel filled with water. In (a) an air bubble remains at the top close to the rubber stopper.
	In addition a water capillary bridge occurs at the rubber O-ring. In (b) the glass barrel is filled with water
	to the rubber O-ring i.e. no air bubble exist.
	}
\end{figure}

\begin{figure}
        \includegraphics[width=0.35\textwidth]{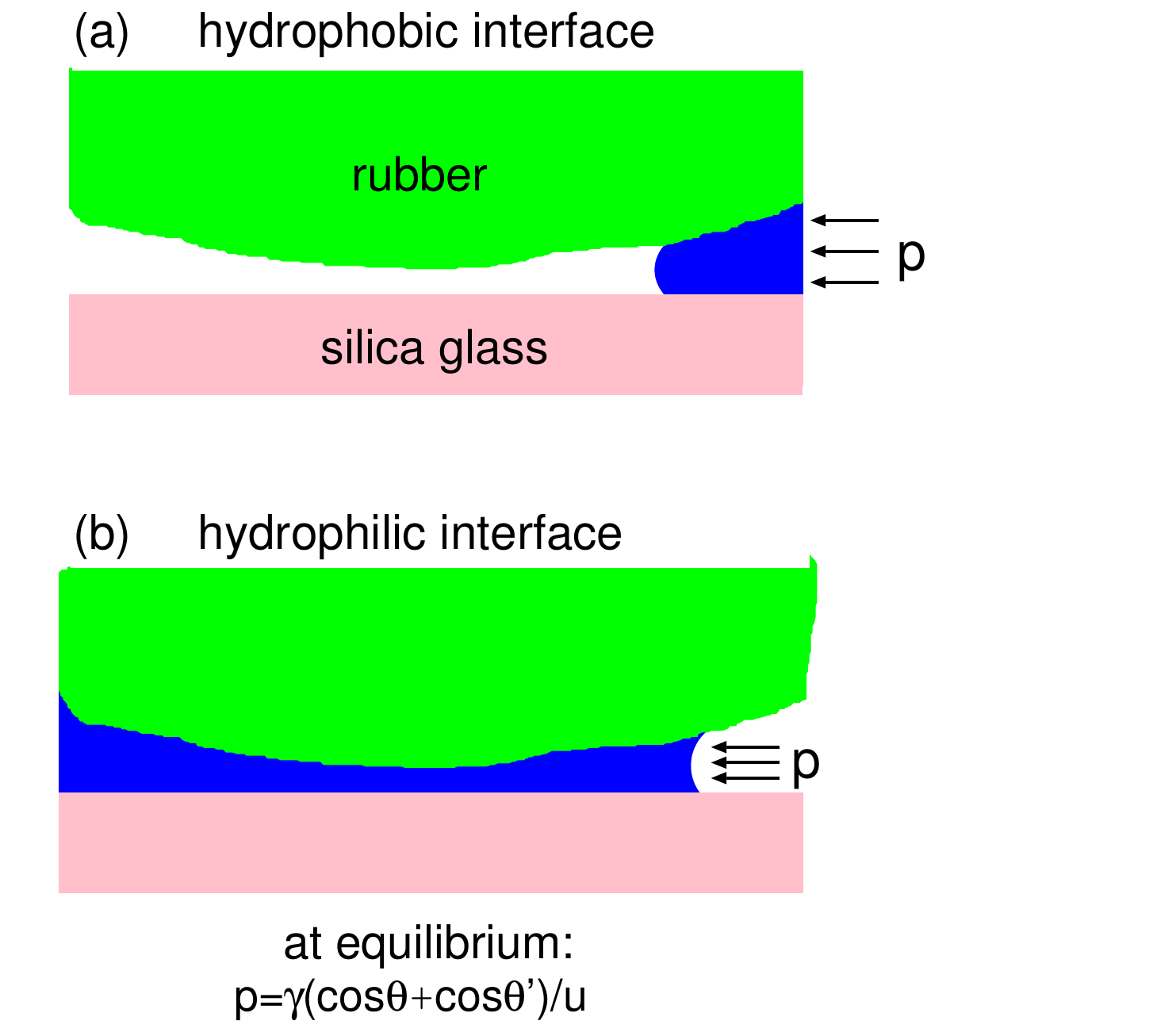}
        \caption{\label{HydrophobicHydrophilic.eps}
	(a) For a hydrophobic interface a finite fluid pressure is needed to squeeze the fluid through the critical junction.
	(b) For a hydrophilic interface a finite fluid pressure is needed to remove a fluid capillary bridge from the critical junction.
	}
\end{figure}

\vskip 0.2cm
{\bf 4.6 Water leakage experiments}

We have performed leakage experiments for syringes filled with water using the J\"ulich 
experimental set-up. We considered two cases where the glass barrel is either completely filled with water
or where an air bubble exist between the inner rubber O-ring and the water. 
        Fig. \ref{1time.2x.mm.O2.second.barrel.in.water.eps}
        shows the displacement of the stopper as a function of time for a water filled glass barrel for O-ring O4.
For $t>5 \ {\rm hour}$ the plunger moves with a nearly constant speed $\approx 1 \ {\rm \mu m/hour}$ corresponding to
a very small leakage rate, $\approx 5\times 10^{-9} \ {\rm cm^3/s}$.
We note that if the air leakage for the unfilled (dry) barrel would be due to viscous flow of the air,
then one would expect the water leakage to be (roughly) smaller than the air leakage 
by a factor of $\approx \eta_{\rm air} / \eta_{\rm water}$ 
where $\eta_{\rm air} \approx 2\times 10^{-5}$
is the viscosity of air and $\eta_{\rm water}\approx 10^{-3} \ {\rm Pas}$ the viscosity of water.
In reality, the air flow is mainly ballistic and the change in the 
leakrate when going from air to water is much larger $\approx 856$
i.e. much bigger than the factor of 50 expected if the air flow would be viscous. The air leakage rate (see Table III) is
$\approx 3.2 \times 10^{-5}  \ {\rm cm^3/s}$ so we expect for water the leakage rate  $\approx (3.2/856) \times 10^{-5} \approx
4\times 10^{-8} \ {\rm cm^3/s}$ which is nearly 10 times bigger than the observed leakage rate. 
In fact the small movement observed (of order $30 \ {\rm \mu m}$ during 40 hours) could be due to viscous relaxation of the rubber
	O-ring due to the axial force, or due to dissolvation of small air bubbles. 

        Fig. \ref{WEToringe.eps} shows the
	O-ring O4 in (a) a dry glass barrel and (b) in a water filled barrel and (c) in a water filled barrel with air bubble
	close to the O-ring. The rubber-glass contact region for the barrels with water must be the same as for the dry
	glass barrel in (a) but due to the influence of the water on the optical picture it appears as if the contact is much
	wider but this is an optical illusion. The important point is that for the water filled barrel with an air bubble, 
	a water capillary bridge occurs between the rubber stopper and the glass barrel resulting in an optical picture very similar
	to the case of a completely water filled barrel.

        Fig. \ref{CapilaryPic.eps}
	shows schematically two barrel filled with water where in (a) an air bubble remains at the top close to the rubber stopper
	while in (b) the glass barrel is completely filled with water. 

	To explain why no (water or air) leakage may occurs for the barrel partly or completely filled with water, note that
        if the interface is hydrophobic in water a finite pressure is needed to squeeze the fluid through the 
	critical junction (see Fig. \ref{HydrophobicHydrophilic.eps}(a)).
	For an hydrophobic interface the (advancing) 
	water contact angles $\theta$ and $\theta'$ on the two solid walls (here rubber and silica glass
	which may be covered by one or a few monolayers of silicon oil) 
        must satisfy ${\rm cos}\theta +{\rm cos}\theta' < 0$.
        The fluid pressure 
	$$p= -({\rm cos}\theta +{\rm cos}\theta'){\gamma \over u}$$ 
	where $u=u(x,y)$ is the surface separation and $\gamma \approx 0.07 \ {\rm J/m^2}$ the water surface tension.
	Using $u=100 \ {\rm nm}$ we get $\gamma /u \approx 10 \ {\rm MPa}$.
	In the present case the water pressure is $\approx 0.1 \ {\rm MPa}$. Hence if $-({\rm cos}\theta +{\rm cos}\theta')>0.01$
	it is impossible to squeeze the fluid through the critical junction, as indeed observed. For an hydrophilic interface
	it may be impossible to remove a fluid capillary bridge but now the resistance to squeeze-out comes from the exit side of the
	critical junction (and involves the receding contact angles)
	rather than on the entrance side as expected for a hydrophobic interface, 
	see Fig. \ref{HydrophobicHydrophilic.eps}(b).

\begin{figure}
	\includegraphics[width=0.3\textwidth]{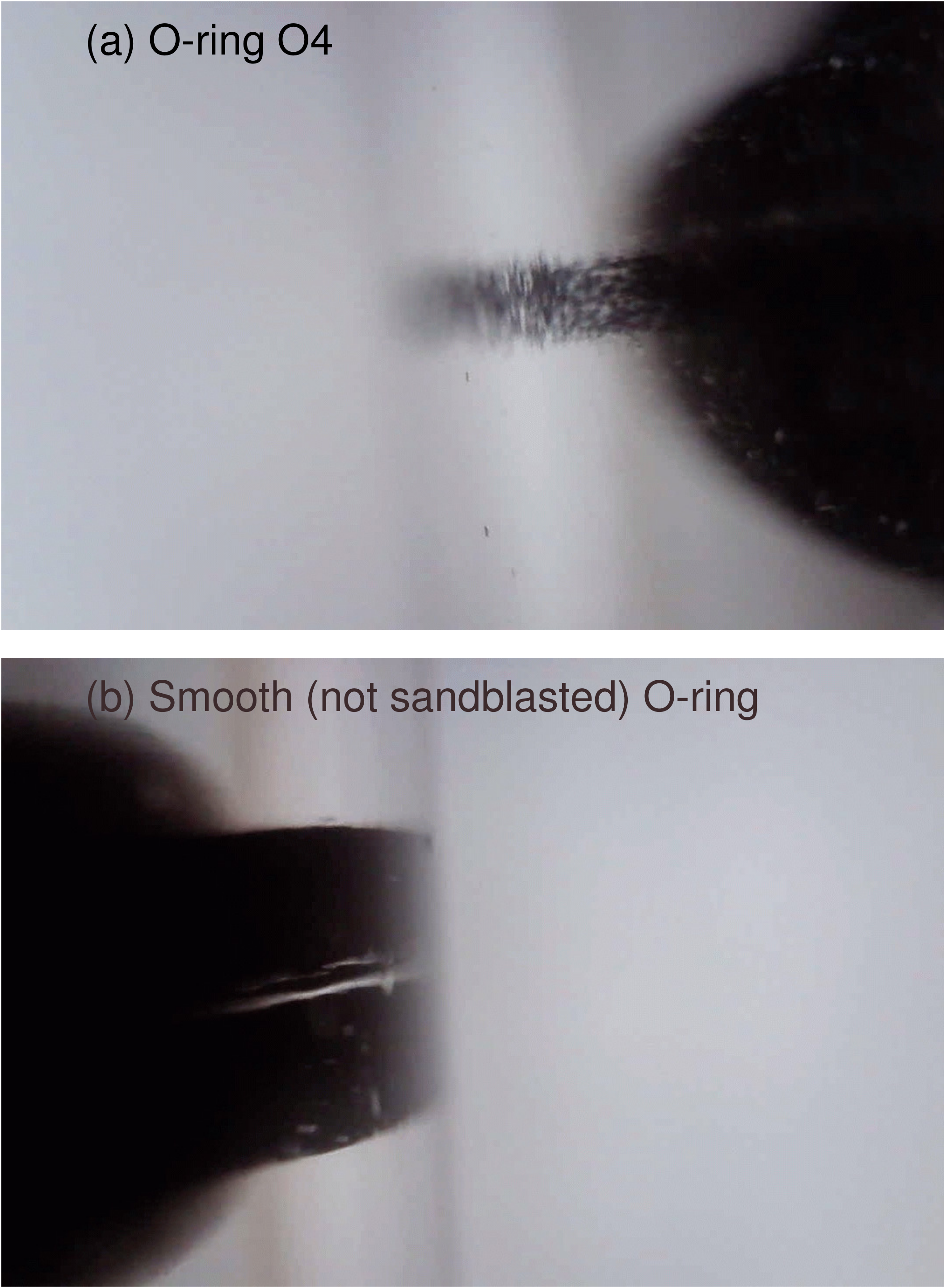}
        \caption{\label{MagnefiedSmooth.andO4.eps}
 The contact region between (a) the sandblasted O-ring O4 and the glass barrel, and
	(b) a not sandblasted  (smooth) O-ring  and the glass barrel. Note the incomplete
	contact in (a) and the apparent complete contact for the smooth O-ring in (b), except for
	a strip (white line) in the middle of the contact due to extra material (flash) 
	formed at the parting lines in the mold	(the flash disappear upon sandblasting).
	}
\end{figure}

\begin{figure}
        \includegraphics[width=0.45\textwidth]{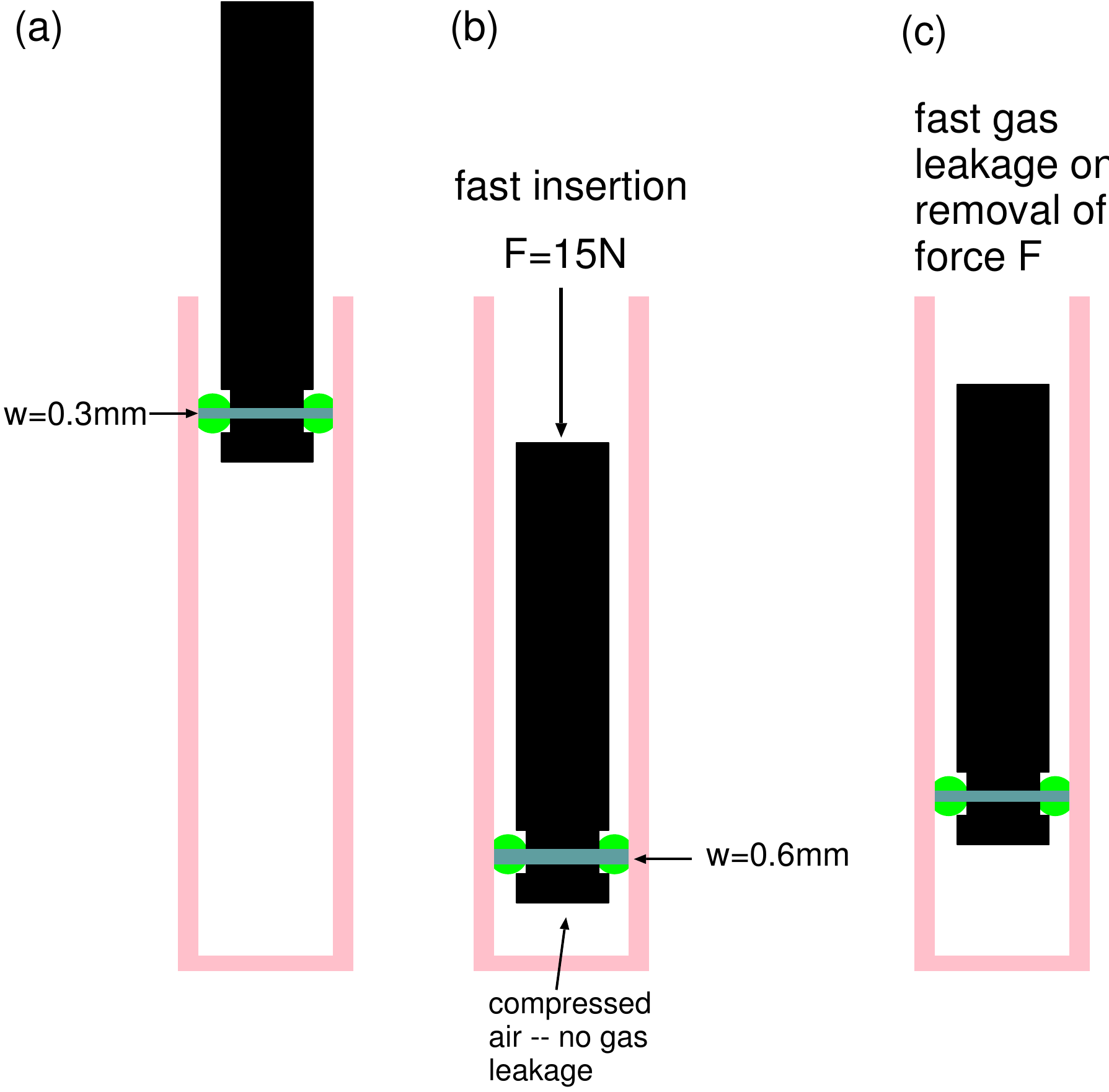}
        \caption{\label{ThreeCasesIlluPic.eps}
(a) When the stopper is inserted slowly in the barrel the rubber-glass contact region has a width
$w\approx 0.3 \ {\rm mm}$. If it  moves slowly towards the top of the barrel the air leak out
and the width of the rubber-glass contact region stays at the value $w\approx 0.3 \ {\rm mm}$
during the whole insertion process. (b) If the stopper is inserted fast, with a big axial force $F\approx 15 \ {\rm N}$,
the rubber-glass contact width increases to $w\approx 0.6 \ {\rm mm}$ while a region of compressed air
forms between the O-ring and the top of the barrel. In this state no air leakage is observed indicating 
that the rubber-barrel contact pressure is so high that the contact area percolate and no open non-contact channels
occur at the rubber-glass interface. (c) If the applied axial force is removes the stopper retract some distance
and the width of the contact region return to its original value $w\approx 0.3 \ {\rm mm}$. 
	If the external force is applied again
the stopper moves much closer to the top of the barrel indicating that air has leaked out during the time period when
the axial force vanished.
}
\end{figure}

\vskip 0.2cm
{\bf 4.7 Discussion}

It would have been interesting to correlate the observed leakrates with optical pictures of the 
rubber-glass contact region. Unfortunately, because of the curved nature of the interface and the
finite resolution of the optical microscopes it is not possible to quantitatively determine the
relative area of real contact $A/A_0$ which would have been of great interest as the theory predict no 
fluid leakage when the area of real contact percolate, which from theory 
(in the absence of adhesion) is expected when $A/A_0\approx 0.42$. Still for the sandblasted O-rings the
optical pictures shows that the contact is incomplete (see Fig. \ref{WEToringe.eps} and \ref{MagnefiedSmooth.andO4.eps}).

When the stopper is pushed into a syringe (with closed needle) compressed air form in the barrel.
Thus the rubber O-ring is exposed to an axial air pressure force in one direction and an opposite
axial force from the plunger rod. As a result they contract a little in the axial direction and since rubber
is nearly incompressible, it tends to expand in the radial direction.
This will increase the rubber-glass contact pressure and hence reduce the leakrate. For the cases studied above
the axial force is relative small and has only a small influence on the leakrate.
However, we performed  experiments with a much bigger axial force, about  $15 \ {\rm N}$, then used before,
and made an interesting observation (see Fig. \ref{ThreeCasesIlluPic.eps}): When inserted in the glass barrel with a small axial force
a sandblasted O-ring had the rubber-glass contact width $\approx 0.3 \ {\rm mm}$.
If the squeezing force is increased to $15 \ {\rm N}$ (giving an air
pressure of about 5 bar) then the width of the rubber-glass contact region becomes $\approx 0.6 \ {\rm mm}$ i.e.
about double of what is obtained when there is no applied squeezing force. This imply that the (maximum) rubber-glass contact 
pressure increases with increasing axial force. In Hertz theory for cylinder against a flat the maximum pressure
$p_{\rm max} = E^*w/(4r_0)$ where $w$ is the width of the contact region and $r_0$ the inner radius of the glass barrel.
Hence increasing the width with a factor of $2$ will increase the maximum pressure by a factor of $2$. In the present case this will
result in a percolation of the contact area in the central area of the rubber-glass contact area. 
Hence no gas leakage is expected
when the axial squeezing force equal $15 \ {\rm N}$. This is in qualitative accordance with our observations. When we remove the
axial force the stopper retract some distance and the width of the rubber-glass contact region (nearly) return to the original
value found when the axial force vanish. At the same time a  ``puff''  sound can be heard. If axial force is immediately
returned to the original value ($15 \ {\rm N}$) the volume of compressed air is much smaller. Clearly, during the short time period where
the axial force was removed, a considerable amount of air was leaking through the rubber O-ring seal. This may in part be due to
the air pressure helping to separate the rubber from the glass in the nominal contact area.

\vskip 0.2cm
{\bf  5. Summary and Conclusions}

A theory was developed for the leakage of gases at the interface between elastic solids with random roughness.  
A simple equation was derived for the gas pressure drop over narrow junctions (critical constrictions),
which takes into account the gas molecule mean free path, 
and the interfacial separation between the solids.
Using this equation and the Bruggeman effective medium theory for the fluid flow conductivity, 
and the Persson's contact mechanics theory for the probability distribution of interfacial separation, 
we predicted gas leak rates in good agreement with experiments.
We applied the theory to gas and water leakage for 
syringes with glass barrels and rubber O-ring stoppers. 
The O-rings were sandblasted to generate well defined random roughness, 
resulting in open (percolating) flow channels in the glass-rubber nominal contact region. 
We studied the leakage of helium, air and water by applying a pressure difference between inside 
and outside the syringe barrel.
For gases (helium and air) the leakage involved ballistic motion of the gas molecules
as mean free path of the gas molecules was of order, or larger than, the mean interfacial separation
in the most narrow junctions along the open flow channels. The 
theory agreed very well with experimental results for leakage of gases.
For water we observed no leakage which we attribute to surface energy effects. 
Thus the applied pressure was not big enough to overcome the Laplace pressure.
We have shown that for (linear) viscoelastic materials the leakrate is independent of the 
effective $E$-modulus because the interfacial separation $u(x,y)$ does
not depend on the $E$-modulus when the compression $\delta$ is given, 
as is the case for a rubber O-ring when confined between the syringe 
plunger road and the barrel.

{\bf  Conflicts of interests}\\
The Authors declare no conflicts of interests among the stakeholders involved in the paper.

\end{document}